\documentclass[aps,prx,twocolumn,superscriptaddress,floatfix,longbibliography]{revtex4-1}
\usepackage{blindtext}
\usepackage{amsmath}
\DeclareMathOperator{\Tr}{Tr}
\usepackage{hyperref}
\hypersetup{
  colorlinks   = true, 
  urlcolor     = blue, 
  linkcolor    = blue, 
  citecolor   = blue 
}

\usepackage{bm}
\usepackage{gensymb}
\usepackage[normalem]{ulem}
\usepackage{euscript}
\usepackage[pdftex]{graphicx}
\usepackage{xspace}
\usepackage{xfrac}
\usepackage{enumerate}
\usepackage{braket}
\usepackage{float,fancyvrb}
\usepackage[T1]{fontenc}
\usepackage{lmodern}
\graphicspath{{./Figures/}}
\usepackage{wrapfig}
\usepackage{scrextend}
\usepackage{comment}
\usepackage{url}
\usepackage{amssymb}
\usepackage{xcolor}

\def\be{\begin{equation}}
\def\ee{\end{equation}}
\def\beq{\begin{eqnarray}}
\def\eeq{\end{eqnarray}}
\def\ba#1{\begin{array}{#1}}
\def\ea{\end{array}}
\def\bn{\begin{enumerate}}
\def\en{\end{enumerate}}

\begin{document}




\title{Thermalization of dilute impurities in one dimensional spin chains}

\author{Dries Sels}
\affiliation{Department of Physics, New York University, New York, NY, USA}
\affiliation{Center for Computational Quantum Physics, Flatiron Institute, New York, NY, USA}
\author{Anatoli Polkovnikov}
\affiliation{Department of Physics, Boston University, Boston, Massachusetts, USA}
\date{\today}

\begin{abstract}
We analyze a crossover between ergodic and non-ergodic regimes in an interacting spin chain with a dilute density of impurities, defined as spins with a strong local field. The dilute limit allows us to unravel some finite size effects and propose a mechanism for the delocalization of these impurities in the thermodynamic limit. In particular we show that impurities will always relax by exchanging energy with the rest of the chain. The relaxation rate only weakly depends on the impurity density and decays exponentially, up to logarithmic corrections, with the field strength. We connect the relaxation to fast operator spreading and show that the same mechanism destabilizes the recursive construction of local integrals of motion at any impurity density. In the high field limit, impurities will appear to be localized, and the system will be non-ergodic, over a wide range of system sizes. However, this is a transient effect and the eventual delocalization can be understood in terms of a flowing localization length.
\end{abstract}

\maketitle

\section{Introduction}

Understanding, and controlling, the conditions under which dynamical systems thermalize under their own internal dynamics is of fundamental interest and has important technological applications. Avoiding thermalization typically requires careful crafting of the Hamiltonian of the system, but it's been proposed that models with local interactions exhibit non-ergodic behavior, that is stable in the thermodynamic limit, when subject to sufficiently large disorder~\cite{basko2006MBL, gornyi2005MBL}. Following these initial publications there has been very extensive work on understanding this non-ergodic phase (currently going by the name of many-body localization (MBL)) and the nature of transition to the ergodic phase. Existence of a well localized regime has been reported by several state of the art experiments~\cite{Schreiber_MBL_2015,bloch2019mbl}. We refer the reader to some recent reviews for further references~\cite{nandkishore2015many, abanin2019review}. Nonetheless, several papers have recently questioned the stability of the MBL phase~\cite{suntajs2019quantum,suntajs2020transition, Kiefer_Emmanouilidis_2021}. In turn the findings of Refs.~\cite{suntajs2019quantum,Kiefer_Emmanouilidis_2021} were challenged by some follow up papers~\cite{abanin2019rebuke, Sierant_2020,LuitzReply,Panda_2020}. Regardless of where one stands in this debate, one of the key challenges in direct numerical, or experimental, study of the MBL transition is that finite size (time) effects at larger disorder are very strong, making it hard to draw unambiguous conclusions. For example, very recently a series of numerical papers, based on newly developed approaches, moved the lower limit of disorder compatible with the MBL transition to much higher values, by factors of two to five more than was previously believed~\cite{morningstar2021avalanches, sels2021markovian,sierant2021observe}.

Early analytical approaches to MBL, starting from the pioneering works~\cite{basko2006MBL,gornyi2005MBL}, focused on the stability of the localized phase against the proliferation of resonances at strong enough disorder, in analogy with the non-interacting problem. The resonances are defined as near degeneracies between localized energy states, which are lifted by the hopping of particles and the interaction between them. It was argued that these resonances cannot destabilize the localized phase at sufficiently strong disorder like in the non-interacting case. A formal mathematical argument for stability of the MBL phase was presented in Ref.~\cite{imbrie2016MBL} for some specific model. In the current work we address the problem from a different angle and come to an opposite conclusion: namely that in the thermodynamic limit the localized phase is always unstable because of off-resonant virtual transitions. While we do not provide a rigorous mathematical proof of this statement, we support our analytical results with a careful numerical analysis. Moreover some of the key numerical results, which were used to demonstrate stability of the localized phase, are consistent with our results.

The approach we develop in this paper is based on first understanding the fate of a single impurity, which is weakly coupled to an ergodic spin chain (bath). At a sufficiently strong local field this impurity undergoes a delocalization crossover as a function of either the bath size $L$ or the impurity observation time $t$: if $L$ or $t$ is small the impurity is effectively localized, only weakly dressed by the bath spins. In the MBL language, it forms a local integral of motion (LIOM)~\cite{abanin2019review}. However, for large $L$ and at sufficiently long times the LIOM decays. We tie this instability to the Krylov complexity of the bath, which was recently proposed as a generic probe of quantum chaos~\cite{parker2018universal,Avdoshkin_2020, Murthy_2019, Cao2021operatorgrowth}. Physically this instability manifests itself in a flowing correlation length $\xi(x)$ with the distance: as the LIOM grows in support, its tails decay slower and slower, leading to eventual divergence of $\xi(x)$. Interestingly we find a direct generic connection between the lifetime of the best (slowest decaying) LIOM and the Fermi Golden Rule (FGR) relaxation rate of the impurity spin. Using this approach we avoid the need of making any assumptions about the structure of the bath eigenstates and can work directly in the thermodynamic limit. 

Having established the connection between the LIOM instability and the FGR rate for a single impurity, we go on to show that the presence of other impurities does not qualitatively change the situation. That is, any finite impurity density will only lead to a finite renormalization of the LIOM relaxation time. The flow and eventual divergence of the correlation length leading to instability of LIOMs is tied to the operator growth, which is not affected by disorder apart from a finite renormalization. The impurity model allows us to carefully study the effect of finite impurity density on the relaxation rate smoothly connecting decay of a single impurity in the presence of disorder with the clean limit. In this way we are able to make predictions about the thermodynamic limit and test them numerically while only using small systems. Such a study are more difficult in canonical MBL models with large random local fields on every site and much larger finite size effects. It is very hard to imagine that there would be any qualitative difference between our model and canonical MBL models. As we discuss later, our findings for the impurity model are in excellent qualitative agreement with both recent and earlier numerical simulations on fully disordered models. Unlike the previous studies we find that the mechanism of delocalization of the impurity in many body systems is not due to proliferation of the resonances but rather due to virtual off-resonant transitions.

While our analytical constructions are rather general, we use a specific model Hamiltonian to support them numerically, namely  
\be
H=H_{\rm bulk}+H_{\rm imp}.
\label{eq:H_total}
\ee
Here 
\be
H_{\rm bulk}= \sum_j ( S^x_j S^x_{j+1}+S^y_j S^y_{j+1}+ S^z_j S^z_{j+1})  + \sum_j h_j S_j^z,
\label{eq:H_XXZ}
\ee
where $S_j^{x,y,z}$ are spin-1/2 operators
is the bulk Hamiltonian describing the bath and 
\be
\label{eq:V_imp_many}
H_{\rm imp}=\sum_j V_j S_j^z,\quad
V_j=\sum_\ell V_\ell \delta_{j \ell},
\ee
where $\{\ell\}$ is a subset of sites where impurities are located and $V_\ell$ are uniformly distributed in the interval $[V/2, 3V/2]$. We allow impurities strengths to fluctuate around mean value of $V$ to avoid dealing with any potential resonances. For one impurity this subset consists of a single site with a fixed strength $V$. The magnetic fields $h_j$ in $H_{\rm bulk}$ are small and random uniformly and independently distributed on all sites in the interval $[-1/4,1/4]$. These magnetic fields serve a two-fold purpose: firstly they break both integrability and translational symmetry of the Heisenberg chain; secondly averaging over disorder allows us to additionally suppress effects of accidental resonances. We checked that all our results reported in this work are valid for each disorder realization. Open boundary conditions are used unless otherwise stated.

We also use the Hamiltonian~\eqref{eq:H_total} to explain how earlier analyses of level statistics is affected by our findings. The absence of scale separation between the freezing of impurities and the decoupling of segments of the chain in small systems, creates the illusion of a fixed crossing but this a purely finite size effect.  Additionally we analyze the fidelity susceptibility for this system recently proposed by us as a probe of chaos~\cite{pandey2020chaos,leblond2020universality} and show that its behavior for a single impurity is very similar to that of a fully disordered model~\cite{sels2020dynamical}. Likewise we find signatures of the inverse frequency scaling of the spectral function ($1/f$-noise) of the bath spins in the presence of the strong impurity, which are also reminiscent of the results found in the fully disordered model~\cite{sels2020dynamical}.

The paper is structured as follows: In the next two sections we analyze the fate of a single impurity, coupled to a weakly disorder chain which serves as a bath, using Fermi's golden rule and a perturbative Birkhoff construction of the LIOM. We explain how these two apparently different approaches are in fact related through the Krylov complexity, and why they lead to the same criterion  for the localization/delocalization crossover at approximately extensive impurity field. The single impurity results not only establish a baseline for understanding how to think about thermalization of the boundaries of rare regions in the putative MBL phase. It also allows us to systematically investigate the affects of additional impurities. We proceed to discuss why adding more impurities to the bath only quantitatively affects the position of this crossover. Finally, we discuss our findings in light of earlier analysis of numerical probes of MBL like level spacing statistics, the fidelity susceptibility and the spectral function of local observables. These probes again highlight qualitative similarity between a single impurity system and fully disordered models. 

\section{Single Impurity}

As a first step in understanding the fate of impurity spins in the Hamiltonian~\eqref{eq:H_total} we consider a setup where a single impurity is weakly coupled to an ergodic bath such that the Hamiltonian is
\begin{multline}
\label{eq:H_bi}
 H_{\rm bi}=H_{\rm bulk}+V S^z_0+\epsilon H_{\rm int}, \\ H_{\rm int}= S^x_1 S^x_0+S^y_1 S^y_0.
\end{multline}
It is convenient to separate the interaction term of the impurity with the bulk into $H_{\rm int}$. The small parameter $\epsilon$ is introduced to control our analytical results. In the numerical analysis of the model we use $\epsilon=1$. As it will become clear shortly the longitudinal coupling $S^z_1 S^z_0$ between the impurity spin and the bath plays no role in our analysis. Formally this term can be always absorbed in $H_{\rm bulk}$ without affecting any results.

\subsection{FGR relaxation}

A standard way to understand relaxation of the impurity coupled to a bath is through the FGR. It is informative to look into the Hamiltonian $H_{\rm bi}$ in the rotating frame defined by the interaction picture of the impurity Hamiltonian $H_0=V S^z_0$, which results into mapping of a static Hamiltonian $H_{\rm bi}$ into a Floquet system with no impurity potential but with a periodically driven hopping between the impurity and the boundary spin:
\be
\label{eq:H_rot_frame}
    H_{\rm bi}^{\rm rot}(t)=H_{\rm bulk}+ 
    \frac{1}{2}\left(e^{-iVt} S^+_{1} S^-_{0}+e^{iVt} S^-_1 S^+_{0} \right). 
\ee
The FGR relaxation rate can be extracted from the spectral function of the oscillating spin-spin coupling in the basis of the bath Hamiltonian. Because the matrix elements of $S_0^{\pm}$ are trivial with respect to $|\!\uparrow\rangle$ and $|\!\downarrow\rangle$ states of the impurity spin it suffices to analyze the spectral function of $S_1^x$ (or equivalently $S_1^y$) of the boundary spin $A_x(\omega)$ defined as:
\be
\label{eq:spectral_function_def}
A_x(\omega)=\int_{-\infty}^\infty \frac{dt}{2\pi}\,\mathrm e^{i\omega t} \mathbb{E}\left[G^n_x(t)\right],
\ee
where $G^n_x$ is the connected correlation function:
\[
G^n_x(t)\equiv 
\frac{1}{2} \langle n| \lbrace S_1^x(t), S_1^x(0)\rbrace_+ |n\rangle_c,
\]
where $\lbrace \dots \rbrace_+$ stands for the anti-commutator.

\begin{figure}[htb]
	\centering
	\includegraphics[width= 0.48\textwidth]{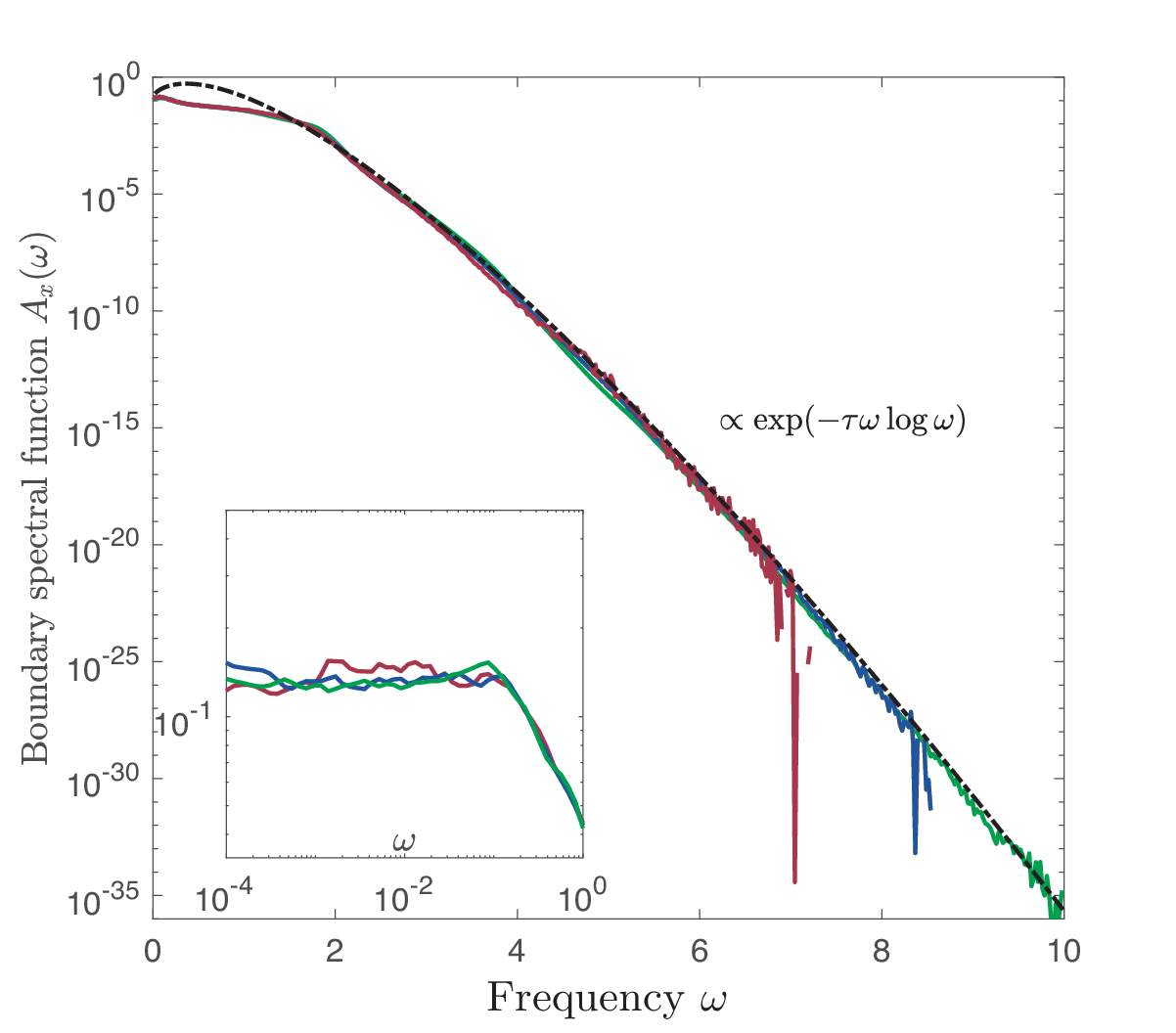}
	\caption{\textbf{Boundary spectrum:} The high frequency part of the spectral function of the $S_1^x$ operator on the boundary of a chain of length $L=12,14,16$ is shown in red, blue and green respectively. The dashed line shows a $A_x(\omega)\propto \exp(-\tau \omega \log(\omega))$ fit, indicating the spectral function saturates the bound expected for 1-dimensional chaotic systems. The inset shows the low frequency part of the spectral function, showing a clear plateau indicative of random matrix theory.}
\label{fig:Xboundary}
\end{figure}

This spectral function is shown at Fig.~\ref{fig:Xboundary} for three different system sizes $L=12,14,16$. At high frequencies the spectral function behaves like 
\begin{equation}
A_x(\omega)\sim \exp(-\tau \omega \log(\omega)), 
\label{eq:speclog}
\end{equation}
where $\tau\approx 3.4$. Note that with increasing system size the spectral function simply extends to higher frequencies. At frequencies below the high frequency cutoff there are almost no finite size effects. This insensitivity of the high frequency response to the system size is consistent with Ref.~\cite{Mukerjee_2006}. The scaling~\eqref{eq:speclog} was predicted earlier as a decay rate of doublons~\cite{Sensarma_doublon_2010, Chudnovskiy_doublon_2012}. It also saturates the upper bound for the spectral function recently derived in Ref.~\cite{Avdoshkin_2020,parker2018universal,Murthy_2019}. In some closely related models the same exponential form can also be shown to be its lower bound~\cite{Cao2021operatorgrowth}. A slightly weaker bound with no $\log(\omega)$ correction was derived earlier in Ref.~\cite{abanin2015heating}. According to the fit shown in Fig.~\ref{fig:Xboundary} this bound is tight and describes the actual spectral function well. In fact this scaling of the spectral function is very easy to understand from simple heuristic considerations. In order to absorb an energy $\omega\gg 1$, the system it is required to use roughly $C \omega$ links as this energy is locally not available. Here $C$ is the constant of the order of one (recall that the spin-spin coupling $J$ on the links is set to unity). Within standard perturbation theory each link will result in $1/\omega$ contribution to the matrix element entering the transition rate, therefore one can estimate the total matrix element as $(1/\omega)^{C\omega}\propto \exp[-C\omega \log(\omega)]$. The square of this matrix element defines the spectral function and correspondingly the FGR decay rate of the impurity spin, which agrees with Eq.~\eqref{eq:speclog} if we identify $\tau=2C$. 

The FGR relaxation rate of a weakly coupled impurity to the boundary inherits the scaling from the spectral function~\cite{Mallayya_2019}:
\begin{equation}
\label{eq:Gamma_FGR}
    \Gamma\propto |\epsilon|^2\exp(-\tau V\log (V\tau)).
\end{equation}
As we already noted the use of FGR is formally justified if we assume that $\epsilon\ll 1$, though it is expected that this relationship between $\Gamma$ and the spectral function holds even when $\epsilon=O(1)$. Note that there is an essential singularity in $\Gamma(V)$ at $1/V\to 0$ such that the relaxation rate cannot be captured in any finite order in perturbation theory in $1/V$. The FGR relaxation should provide an effective mechanism for the impurity spin relaxation as long as it is much larger than the level spacing: $\Delta=\exp[-S(L)]$. The criterion is equivalent to demanding that the typical unperturbed susceptbility $\chi$ for switching on the coupling between the impurity spin and the rest of the chain would be larger than $O(1)$, i.e. that the eigenstates of the impurity and the bath will fully mix with each other, see Appendix A. We thus conclude that the critical impurity potential separating the localized and delocalized regime scales as
\begin{equation}
\label{eq:V_FGR}
V^\ast(L)\approx {S(L)\over \tau \log(S(L)/\tau)}\sim {L \log(2)\over \tau \log(L \log(2)/\tau)}
\end{equation}
Up to the logarithmic correction the critical impurity potential separating localized and delocalized regimes scales linearly with the bath size.

\subsection{Asymptotic Birkhoff construction of the LIOM}
In small systems, where $\Gamma\ll \Delta$ FGR does not apply. Instead it is expected that the boundary spin will only partially relax and form a so called local integral of motion (LIOM)~\cite{abanin2019review}. To test this idea, we will construct the LIOM in the leading order of perturbation theory in the coupling to the bath $\epsilon$ and in all orders in $1/V$. This is exactly the same order of approximation which is used to derive the FGR. To construct the LIOM we use the so called Birkhoff normal form, where we build a conserved charge iteratively as a series in $1/V$:
\begin{equation}
\label{eq:expansion_tau_z}
    Q=S^z_0+{1\over V} q_1(\epsilon)+{1\over V^2} q_2(\epsilon) +\dots,
\end{equation}
requiring that in each order in $1/V$ the commutator $[Q,H]$ vanishes to the same order in $1/V$.

This equation can be solved order by order. Using that $[S_z^0, H_{\rm bulk}]=0$ and $\{S_z^0, H_{\rm int}\}_+=0$ it is easy to check (see also Appendix B) that to linear order in $\epsilon$ and $n$-th order in $1/V$ the LIOM is given by
\begin{multline}
\label{eq:nested_commutator_Q}
    Q_n=S^{z}_0+\epsilon \sum_{q=0}^n {1\over V^{2q+1}} {\rm Ad}_{H_{\rm bulk}}^{\,2q} H_{\rm int}\\+\epsilon \sum_{q=1}^n {1\over V^{2q}} [{\rm Ad}_{H_{\rm bulk}}^{\,2q-1}H_{\rm int}, S_0^z] 
\end{multline}
The norm of nested commutators entering the expansion  $R_k\equiv i^k {\rm Ad}_{\rm H_{\rm bulk}}^{\, k} H_{\rm int}$ is tied to the parameter $\tau$ defining the FGR decay rate. Namely,
\[
\|R_k\|^2\equiv  {1\over 2^L} {\rm Tr } (R_k^2),
\]
where $L$ is the system size. At large $k$ this asymptotes to~\cite{parker2018universal, Avdoshkin_2020,Cao2021operatorgrowth}:
\begin{equation}
\label{eq:norm_nested_avdoshkin}
    \|R_{k}\|^2\sim \left({2k\over {\rm e}\,\tau  \ln(2k)}\right)^{2k}.
\end{equation}
Using cyclic properties of the trace it is easy to check that for any integers $k$ and $q$ we have ${\rm Tr}[R_{k}R_{k+2q+1}]=0$ and ${\rm Tr}[R_{k}R_{k+2q}]={\rm Tr} [R_{k+q}^2]$. 
This observation allows us to exactly account for the interference between different terms in the expansion and express the norm of the conserved operator through the sum of norms of operators $R_k$ with positive coefficients:
\begin{eqnarray}
\label{eq:norm_Qn}
   && ||Q_n||^2\equiv 1+\epsilon^2 \sum_{k= 1}^{n} C_k^{(n)} {||R_k||^2\over V^{2k}}, \\
   && C_k^{(n)}=\left\{ 
        \begin{array}{ll}
          2k-1 & k<n/2\\
          2(n-k)+1 & k\geq n/2
         \end{array}
    \right. .\nonumber
\end{eqnarray}
The norm of the residual of the commutator of $[Q_n,H]$ determines the lifetime of the operator $Q_n$ as it follows from the short time expansion of the nonequal time correlation function ${\rm Tr}\,[ Q_n(t)Q_n(0)]$~\cite{kim_2015_slowest}:
\begin{multline}
\label{eq:Gamma_N}
    \Gamma_n^2=\|i [Q_n,H]\|^2\approx \epsilon^2 {\|R_{2n+1}\|^2\over V^{4n+2}}\\
    \sim \epsilon^2 \left({4n+2\over {\rm e}\, V \tau  \ln(4n+2)}\right)^{4n+2}
\end{multline}
Expression~\eqref{eq:norm_nested_avdoshkin} makes clear that the Birkhoff construction is asymptotic. At large $V$, the decay rate has a non-monotonic dependence on $n$. It is convenient to introduce the running localization length as
\begin{multline}
\xi(n)=-\left({d\log \Gamma_n\over dn}\right)^{-1}\\
\approx {1\over 2 \left(\log (V\tau)+\log\log(4n+2)-\log(4n+2)\right)}
\end{multline}
This localization length flows with $n$ diverging at
\[
n=n^\ast\approx {V\tau \over 4} \log(V\tau).
\]
At this value of $n^\ast$ the perturbative decay rate $\Gamma_n^2$ reaches its minimum
\begin{equation}
\label{eq:Gamma_min}
    \Gamma_{n^\ast}^2=\Gamma^2_{\rm min}\approx \epsilon^2 \exp[-V\tau \log(V\tau)].
\end{equation}
Apart from an overall prefactor the square of the short time decay rate of the LIOM: $\Gamma_{\rm min}^2$ coincides with the FGR rate $\Gamma$~\eqref{eq:Gamma_FGR}. This situation is not unexpected and a complementary discussion can be found in Ref.\cite{morningstar2021avalanches,sels2021markovian}.

Physically for local Hamiltonians the index $n$ represents the spatial range of the approximate LIOM $Q_n$. The flow of the localization length $\xi(n)$ for $n<n^\ast$ indicates that the decay of the tails of this LIOM slows down with the distance. Eventually the decay stops when the localization length diverges. 


\subsection{Variational conserved charge}
One might wonder if this divergence can be regularized in some way leading to a better conserved charge $Q_n$. To address this question we can use a variational approach using the same commutator ansatz as in perturbation theory as a basis, but allowing for arbitrary coefficients. Instead of computing the coefficients in front of nested commutators in Eq.~\eqref{eq:nested_commutator_Q} perturbatively we will assume that they are variational parameters. It is easy to check that in the limit $n\to \infty$ this variational ansatz is exact in the linear order in $\epsilon$. The contributions can be generated recursively as follows. Given an operator $O_n$ at order $n$, define
\begin{equation}
    O_{n+1}=[S^z_0,[H_{\rm bulk},O_n]].
\end{equation}
The variational ansatz thus consists of an arbitrary operator in the Krylov subspace of the superoperator $\mathcal{B}(\cdot)=[S^z_0,[H_{\rm bulk},\cdot]]$. It's insightful, and indispensable to perform the optimization numerically, to Gram-Schmidt orthogonalize the basis operators as they are generated. The procedure is similar to the familiar Lanczos procedure to generate Krylov space of the Liouvillian~\cite{parker2018universal}. Given a charge $q_n$ at order $n$, define
\begin{equation}
    p_{n+1}=[S^z_0,[H_{\rm bulk},q_n]],
\end{equation}
which is proportional to the next order term. To generate an orthonormal set it suffices to orthogonalize it with respect to the last two $q_n$. Hence we define:
\begin{equation}
q_{n+1}=\gamma_n \left( p_{n+1}-\alpha_n q_n -\beta_n q_{n-1} \right),
\end{equation}
where 
\begin{eqnarray}
    \alpha_n=\Tr [q_n^\dagger p_{n+1}], \\
    \beta_n=\Tr [q_{n-1}^\dagger p_{n+1}], \\
    \gamma_n^{-2}= \Tr [p^\dagger_{n+1}p_{n+1}]-\alpha_n^2-\beta_n^2,
\end{eqnarray}
which makes the charges obey 
\begin{equation}
    \Tr[q_n q_m]=\delta_{n,m}.
\end{equation}
To generate the same set of operators as the Birkhoff construction from the previous section we use $q_0 \propto H_{\rm int}$.
Then we can write the variational conserved charge as 
\begin{equation}
    Q_n^{\rm var}= S^z_0 + \epsilon \sum_{k=0}^n \psi_k q_k.
\end{equation}
The best variational solution could be defined as the one which minimizes the residual commutator with the Hamiltonian, i.e. 
\begin{equation}
\Gamma^2_{\rm opt}= \underset{\psi}{\rm argmin} \| [Q_n^{var},H_{\rm bi}]\|^2.
\end{equation}
In the leading order in $\epsilon$ this  yields the set of linear equations for $\psi$: $S\psi=f$ with the matrix $S$ having elements
\begin{equation}
 S_{km}={1\over 2^L}\Tr[[H_0,q_k][q_m,H_0]] ,\quad H_0=V S^z_0+H_{\rm bulk} 
\end{equation}
and the source vector $f$ defined as
\begin{equation}
f_k=\Tr[[H_0,q_k][q_0,S^z_0]].
\end{equation}
Using the basic definitions of $q_k$ and the fact that $[S^z_0,[S^z_0,q_k]]=q_k$ it's rather straightforward to show that $S$ is a real symmetric pentadiagonal matrix. In addition, the source term $f$ is only non-zero at the first two entries $k=0,1$. 
\begin{figure}[htb]
	\centering
	\includegraphics[width= 0.48\textwidth]{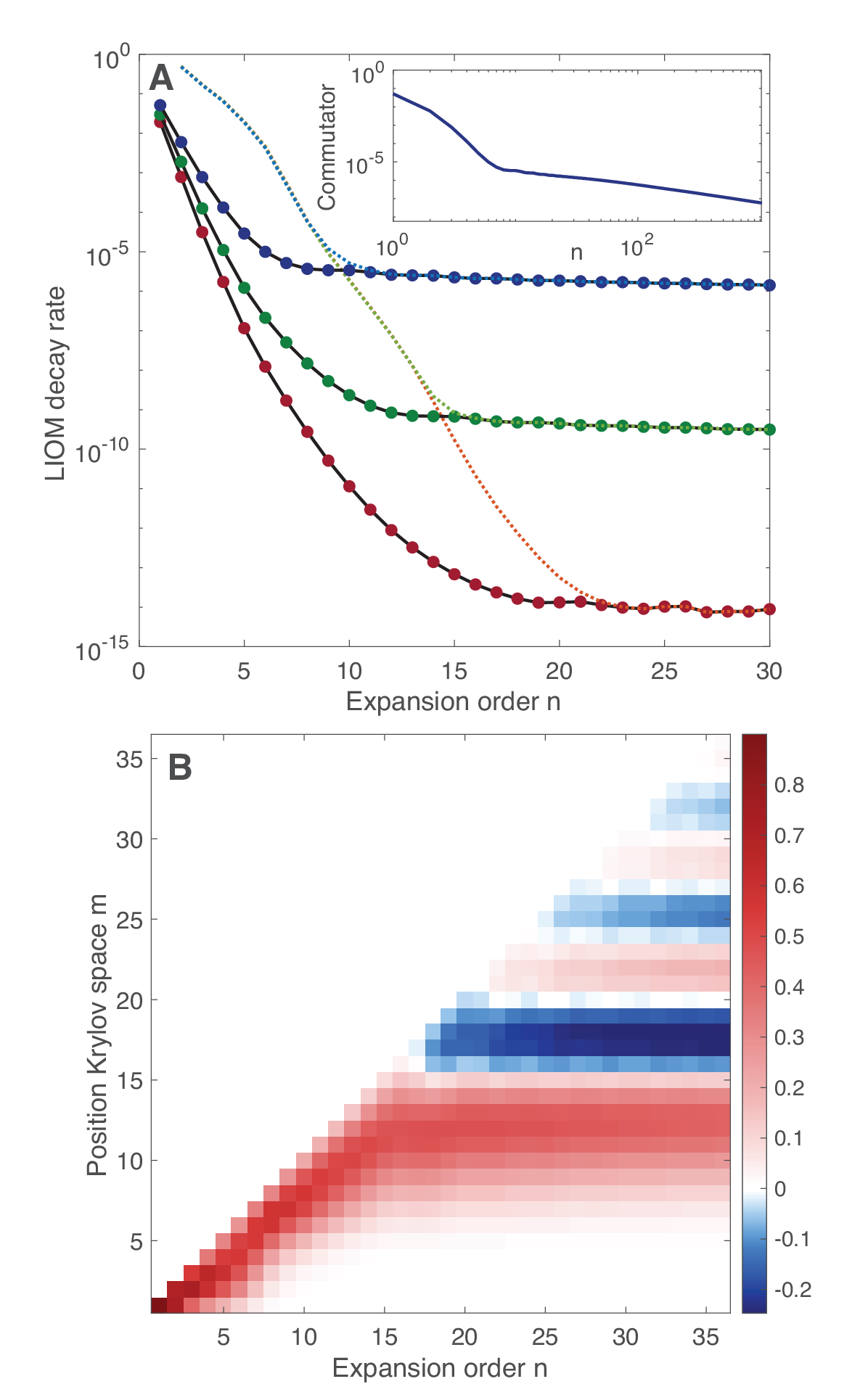}
	\caption{\textbf{Variational LIOM decay:} (Panel A) Residual decay rate of the $n^{\rm th}$ order variational Birkhoff approximation of the integral of motion associated with the impurity spin.  The full circles show the best variational approximation composed out of all operators to order $n$. The dotted lines show the results obtained when the operators are constrained to all but the lowest eigenvalue eigenmodes of $S$. They become nearly degenerate at the crossover. Different colors show different values of the impurity potential, i.e. blue, green, red correspond to $V=3,4,5$ respectively. The inset show $V=2$ up to order $n=1000$. (Panel B) Ground state wave function on the Krylov space for $V=4$ (green curve in panel A), showing a drastic change in behavior at the crossover.}
\label{fig:birkhoffvar}
\end{figure}Further analytical progress seems possible, but we postpone that to future work. We only point that in the limit of large V we can easily recover the perturbative solution, see Appendix C for more details. To proceed here we solve the problem numerically, and the results are summarized in Fig.~\ref{fig:birkhoffvar}. We make a number of observations, first at low order $n$ the variational results agree with the perturbative construction from the previous section. At the crossover $n^\ast$, where the perturbative result yields the minimal relaxation rate, the variational improvement stalls and the residual reaches a plateau. Second, since $S$ is a Hermitian matrix the solution can be decomposed in the eigenmodes of $S$ and we can investigate the stability of the problem by removing the most irrelevant modes one by one. For low orders $n<n^\ast$, we observe a a drastic increase in the relaxation meaning that the best mode is well defined. At $n\approx n^\ast$ this gap closes. In Fig.~\ref{fig:birkhoffvar}, we also observe a clear transition in the ground state of $S$ from localized near the diagonal, to oscillatory with it's mode fixed near $n^\ast$.

From this analysis we conclude that in finite size systems, as long as $L \lesssim n^\ast\sim V\tau \log(V\tau)$, there is a well defined LIOM, adiabatically connected to the boundary spin operator $S_0^z$. For larger system sizes this LIOM becomes unstable and delocalizes. In this sense the LIOM is similar to a long lived quasi-particle, which eventually decays.
Two seemingly different criteria for localization-delocalization crossover: i) FGR rate becomes of the order of the level spacing and ii) Birkhoff LIOM construction starts to break down, thus lead to the same estimate of the bath size corresponding to this crossover. This agreement between the two approaches is not accidental as both results are ultimately connected to the universal operator growth of the nested commutators of $H_{\rm int}$ and $H_{\rm bulk}$. It also substantiates the idea that once the recursive Birkhoff construction breaks down, the system starts thermalizing; eventually becoming ergodic. In that sense the present work is entirely along the lines of seminal works by Abou-Chacra,Thouless and Anderson~\cite{Abou_Chacra_1973} and Basko, Aleiner and Altshuler~\cite{basko2006MBL}, where the authors construct a self-consistent theory of localization, solve the equations order by order and interpret the instability of the construction as a sign of delocalization. While our construction is different, we establish a more direct link between both sides of the transition. Finally, in Appendix D we present a brief discussion on how the same construction can be applied to periodically driven systems, where instead of an impurity one can couple a harmonic oscillator to a spin chain. We tie heating to the divergence of the LIOM connected to the "photon" number.

\section{Finite Impurity Density}
Now let us see how the previous analysis is modified if we consider the full Hamiltonian~\eqref{eq:H_total} with a finite density of impurities. We will still use the impurity at the edge as a probe, i.e. analyze the Hamiltonian~\eqref{eq:H_bi}, where $H_{\rm bulk}\to H=H_{\rm bulk}+H_{\rm imp}$. Clearly a straightforward application of the Birkhoff construction fails as $H_{\rm bulk}$ contains terms of the order of the impurity potential $V$ such that the expansion~\eqref{eq:expansion_tau_z} becomes much more complicated, e.g. the probe impurity could resonate with some other impurity which could lead to an instability in the naive Birkhoff construction that would not necessarily imply delocalization. To tackle this problem we will first perform a Schrieffer-Wolff (SW) transformation on the bath Hamiltonian to effectively eliminate the impurity spins. In particular, if the bath contains a single impurity at a site $\ell$ then after the SW transformation we obtain the following effective Hamiltonian describing the bath
\begin{multline}
 \label{eq:HSW}
    \tilde{H}_{\rm SW}=\left[H_L+S^z_{\ell-1} \left(\frac{1}{4V_{\ell}}+\Delta S^z_{\ell} \right)\right]\\
    +\left[H_R+S^z_{\ell+1} \left(\frac{1}{4V_{ \ell}}+\Delta S^z_{\ell} \right)\right]  \\
    +\left(V_{\ell}-\frac{1}{2V_{\ell}}\right) S^z_{\ell}
    +\frac{1}{V_\ell} S^z_{\ell} \left( S^x_{\ell-1} S^x_{\ell+1}+S^y_{\ell-1} S^y_{\ell+1}, \right),
\end{multline}
where $H_L$ and $H_R$ describe the blocks of the bath Hamiltonian on the left and on the right of the site $\ell$. By construction, the Hamiltonian is still diagonal in the bath-impurity spin and one can thus consider the two sectors with $S^z_\ell=\pm 1/2$ independently. 

This transformed Hamiltonian can be obtained in two different ways: i) either by performing a standard unitary rotation, which perturbatively removes the coupling between the impurity and the rest of the spins:
\begin{equation}
\tilde{H}_{\rm SW}=e^{iK} H e^{-iK}, 
\label{eq:SWtransform}
\end{equation}
where
\begin{multline}
    K= S^y_{\ell} \left( \frac{1}{V_\ell}\left(S^x_{\ell-1}+S^x_{\ell+1} \right)-\frac{i}{V^2} [\left(S^y_{\ell-1}+S^y_{\ell+1} \right),H]\right) \\
    + S^x_{\ell}\left( \frac{-1}{V_\ell} \left(S^y_{\ell-1}+S^y_{\ell+1} \right)- \frac{i}{V_\ell^2} [\left( S^x_{\ell-1}+S^x_{\ell+1}\right),H] \right)\\
    +O(V_\ell^{-3}).
\end{multline}
or ii) by going to the rotating frame with respect to the impurity potential like in Eq.~\eqref{eq:H_rot_frame} and performing the leading order van Vleck expansion of the resulting Floquet Hamiltonian~\cite{Bukov_SW_2016}. 

The transformed Hamiltonian $\tilde H_{\rm SW}$ in Eq.~\eqref{eq:HSW} is quite simple, apart from some boundary corrections arising from the bare interaction and virtual coupling with the impurity, it has some effective flip-flop contribution coupling the boundary spins of the two blocks. Since that coupling arises from a virtual process involving the impurity, it's suppressed by $1/V_\ell$. Note that the sign of the coupling, determined by the value of $S^z_\ell$, is irrelevant as it can be removed by a simple $\pi$-rotation of spins on the right of the impurity around the z-axis. We restrict ourselves to $S^z_\ell=1/2$ sector in Eq.~\eqref{eq:HSW}. The other two $1/V_\ell$ corrections to the transformed Hamiltonian, representing small boundary magnetic fields, are unimportant for the physics of the model and we drop them, defining the effective bulk Hamiltonian: 
\begin{multline}
H'_{\rm bulk}=H_L+H_R+\frac{\Delta}{2}(S^z_{\ell-1}+S^z_{\ell+1})\\
+\frac{1}{2V_\ell} \left( S^x_{\ell-1} S^x_{\ell+1}+S^y_{\ell-1} S^y_{\ell+1} \right).
\label{eq:Heff}
\end{multline}
If the bath contains multiple impurities this procedure can be done on each impurity site effectively introducing weak links across them. The new bulk Hamiltonian $H'_{\rm bulk}$ contains no large terms of the order of $V$ and one can thus apply both the FGR and the Birkhoff analysis by simply replacing the bulk Hamiltonian $H_{\rm bulk}\to H'_{\rm bulk}$ in Eq.~\eqref{eq:H_bi}.

First, we want to understand how the presence of multiple impurities affects the spectrum of the boundary spins shown  in Fig.~\ref{fig:Xboundary}. A sufficient condition for localization is vanishing of the spectral function at $\omega\geq V$ for a sufficiently large $V$. This would ensure that the FGR rate for impurity relaxation is zero. Conversely finite spectral weight at any (non-extensive) $V$ indicates delocalization of the impurities. Clearly, when focusing on a single impurity, the rest of the system acts as the worst bath when all the other impurities are completely frozen out. It thus suffices to understand the modifications of the boundary spectral function due to the presence of additional weak links in the bulk of the bath. 
\begin{figure}[ht]
	\centering
	\includegraphics[width= 0.48\textwidth]{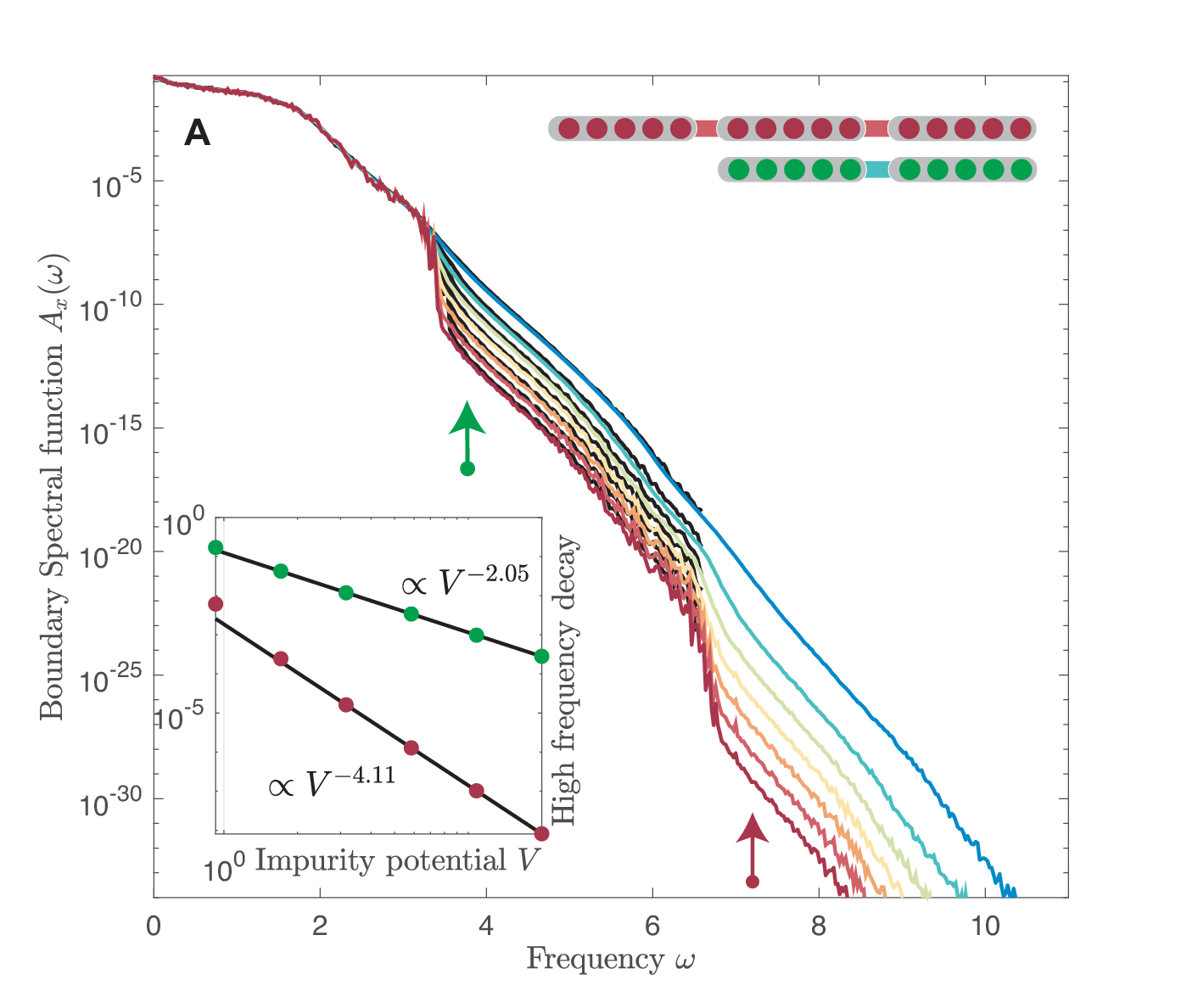}
    \includegraphics[width= 0.48\textwidth]{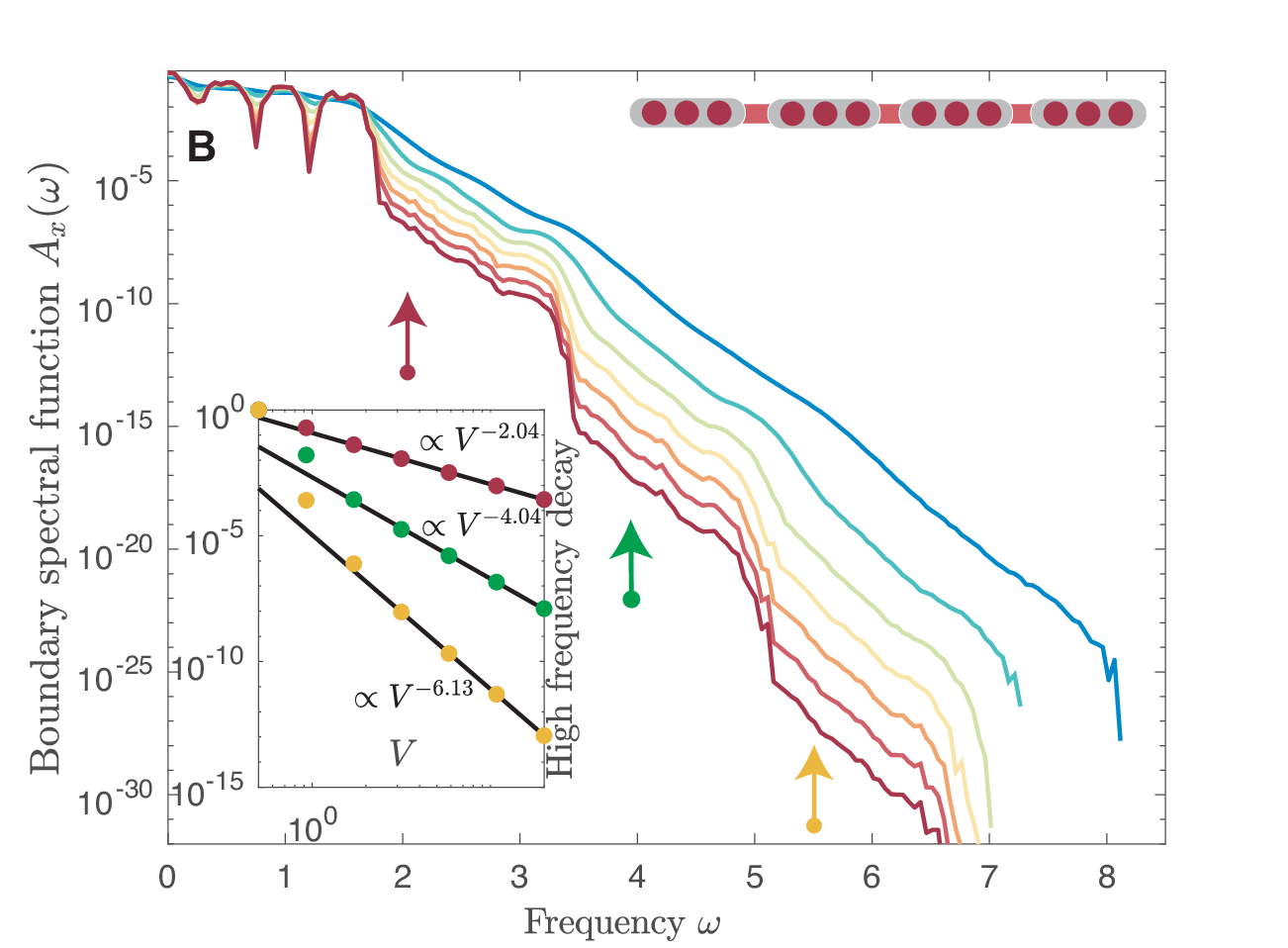}
	\caption{\textbf{Boundary spectrum II:} The high frequency part of the spectral function of the $S^x$ operator on the boundary of a chain with a weak link $J^{\rm eff}_\ell=1/(2V_\ell)$ after  every fifth site (panel A) and every third site (panel B); color goes from blue to red with increasing $V$. The black dashed lines in panel A are for a system of $L=10$, indicating that the intermediate frequency part remains unchanged with increasing system size. The insets show the scaling of the jumps in the spectral function at the frequencies indicated by the corresponding arrows in the main figure. }
\label{fig:BoundaryspecWeaklink}
\end{figure}
The corresponding spectral functions for the boundary spin $S_x^{1}$ are shown in Fig.~\ref{fig:BoundaryspecWeaklink}
for two different arrangements of weak links corresponding to $\ell=\{6,12\}$ and $\ell=\{4,8,12\}$ in expression~\eqref{eq:V_imp_many}. The top panel shows the results for the effective model of size $L=10$ with one weak link in the middle and $L=15$ with two weak links. Different colors correspond to different impurity potentials ranging from 1/2 to 20, specifically $V= 40^{k/6}/2$, $k=0,1,\dots 6$, and hence different strengths of weak links $J^{\rm eff}_{\ell}=1/(2V_\ell)$ (recall that $V_\ell\in [V/2, 3V/2]$). Like in Fig.~\ref{fig:Xboundary} the spectral functions for different system sizes look identical up to the cutoff scale which increases with the many-body bandwidth. Compared to the case with no impurities, which also corresponds to the top blue line corresponding to $V=1/2$, we see two jumps developing in the spectral function at $\omega\approx 3.5$ and $\omega\approx 7$. These jumps can be easily explained using the same heuristic argument as before: In order to dump a large amount of energy $\omega$ one has to excite $\tau\omega/2$ strong links (see discussion after Eq.~\eqref{eq:speclog}). However, after each $\Delta\ell=5$ strong links in our setup there is a weak link, which almost does not contribute to the energy if $J^{\rm eff}_\ell\ll 1$ but leads to an additional $1/(2V_\ell)$ suppression to the matrix element and correspondingly $1/(2V_\ell)^2$ suppression to the spectral function and the FGR rate. This simple argument is confirmed numerically in the inset of the top panel Fig.~\ref{fig:BoundaryspecWeaklink} where the two lines show dependence of the drop in the spectral function on $V$ at two values of $\omega$ indicated by the arrows in the main plot. The extracted jumps are well described by power laws consistent with the expected $1/(2V)^2$ (after one jump) and $1/(2V)^4$ (after two jumps) scalings. In the bottom panel of Fig.~\ref{fig:BoundaryspecWeaklink} we show similar results for weak links located after every third site as shown in the inset. Now the jumps appear more frequently but the magnitude of each jump is again consistent with $V^{-2}$ scaling per block. We thus see that the spectral function of the model with weak links is described by
\begin{equation}
A^{\rm eff}_x(\omega) 
\gtrsim A_x(\omega) \exp\left(-{\tau \omega\over \Delta\ell} \log(2V)\right).
\end{equation}
This spectral function gives a lower bound on the spectral function of the full model (see Appendix E) and hence defines a lower bound on the FGR relaxation rate of the impurity
\begin{multline}
\Gamma\geq A^{\rm eff}_x(V)\sim \mathrm e^{-\tau' V\log(V\tau')}=\Gamma_0^{1+1/\Delta\ell},\\ \tau'=\tau(1+1/ \Delta\ell)
\end{multline}
We thus conclude that the lower bound of the FGR decay rate of the impurity is only weakly affected by the presence of other impurities, which somewhat increase the effective exponent $\tau\to \tau'$. As a consequence, for any impurity at a finite energy $V$, or more accurately at $V$ which increases slower with system size than $L/\log(L)$, there is a sufficient spectral weight to dissipate energy into the bath.

Like in the single impurity case one can check stability of LIOMs when the FGR relaxation rate becomes smaller than the level spacing. The Birkhoff perturbative construction of the LIOM associated with the boundary impurity looks the same as in the single impurity case with the only difference that we will encounter a finite density of weak links in the nested commutators $R_k$ appearing in Eq.~\eqref{eq:norm_Qn} such that the norms of such commutators will be suppressed by at most $V^{2k/\Delta\ell}$ if we assume that weak links appear in the rate $1/\Delta\ell$. Suppression is likely even less as the norm will be dominated by the terms containing fewer than average weak links. In either case this suppression is not enough to counter the factorial growth of the norms nested commutators $\|R_k\|$. Moreover Refs.~\cite{Ballar_Trigueros_2022} and~\cite{trigueros2021krylov} argued that for fully disordered models the asymptotic behavior of these norms at large $k$ is not affected by the disorder potential except for finite renormalization of the parameter $\tau$. As a result the LIOM associated with the probe impurity spin remains perturbatively unstable for any density of weak links. 

\begin{figure}[htb]
	\centering
	\includegraphics[width= 0.48\textwidth]{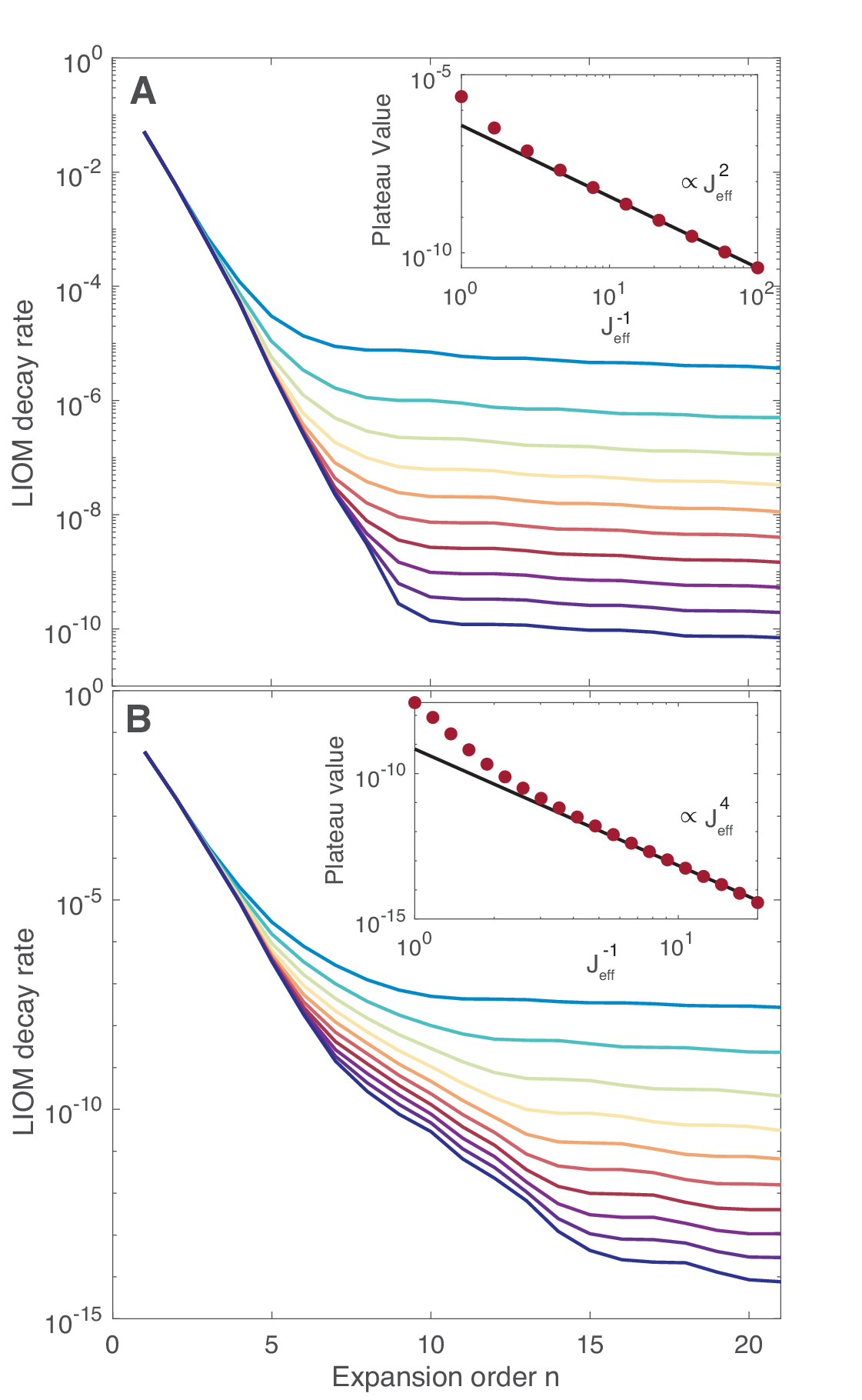}
	\caption{\textbf{Variational LIOM decay II:} (Panel A) Residual decay rate of the $n^{\rm th}$ order variational Birkhoff approximation of the integral of motion associated with the impurity spin. The chain has a weak link $J_{ \rm eff}$ after every third site and the external impurity field is $V=3$; color goes from blue to red with decreasing $J_{\rm eff}$ ranging from 1 to 100. The inset shows the plateau value scales like $J_{\rm eff}^2$. (Panel B) Shows the same decay rate for an impurity field of $V=4$, with the inset highlighting the plateau value now scales as $J_{\rm eff}^{4}$, ranging from 1 to 20.}
\label{fig:birkhofflinks}
\end{figure}

\begin{figure}[htb]
	\centering
	\includegraphics[width= 0.48\textwidth]{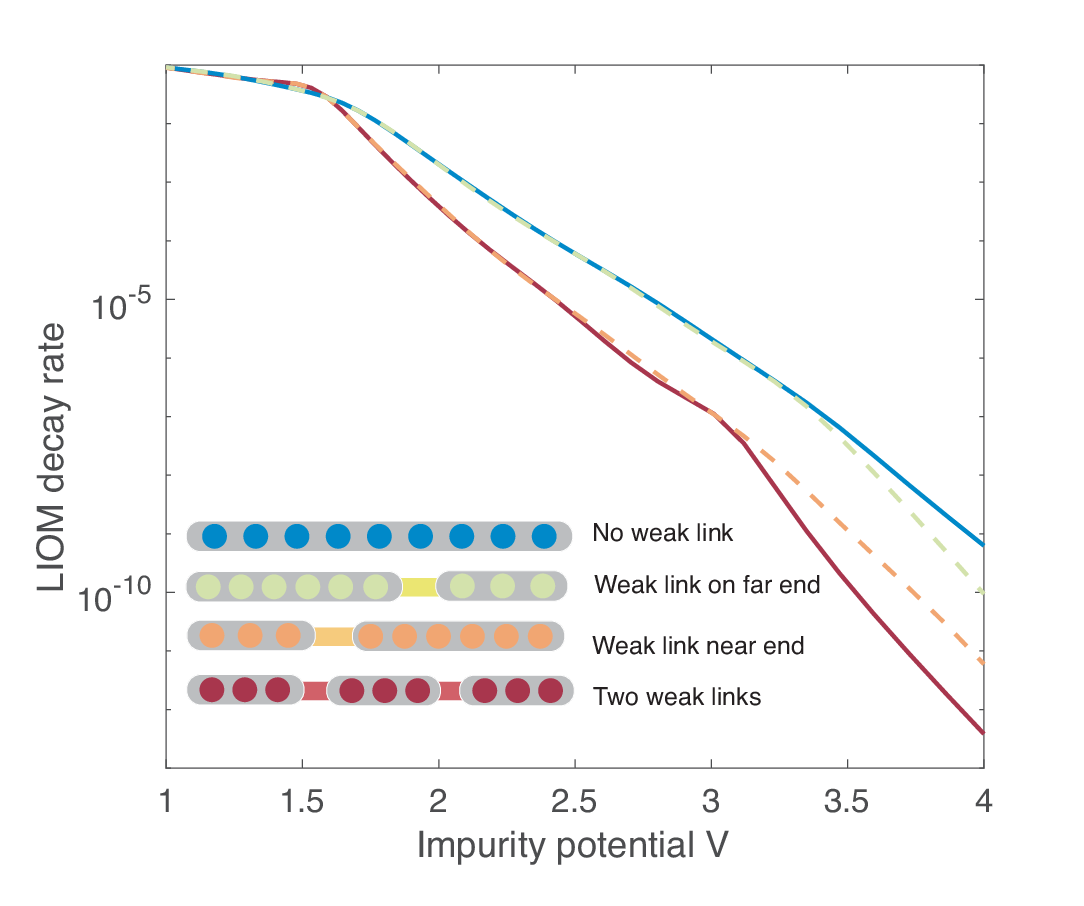}
	\caption{\textbf{Self consistent rate:} Residual decay rate of the asymptotic variational Birkhoff approximation of the integral of motion associated with the impurity spin, connected to a  chain with weak links as a function of the impurity potential $V$. The weak links are chosen self-consistently like $J_{\rm eff}=1/2V$. The blue line shows the results with no weak links, the red with weak links after every third site, the dashed orange line would be the result if only the weak links closest to the impurity is taken into account and the dashed green line if only the second closest weak link were to be there. The results highlight that the decay is only suppressed by weak links that appear at a distance before the crossover scale $n^\ast$.}
\label{fig:birkhofflinksselfconsistent}
\end{figure}

To confirm this, we construct the same variational LIOMs as for the single impurity problem within the effective weak link model. The results are summarized in Fig.~\ref{fig:birkhofflinks}, where we observe suppression of the residual commutator with $J_{\rm eff}^2$ when $V=3$ and $J_{\rm eff}^{4}$ for $V=4$. By increasing the impurity potential from $3$ to $4$ we increase $n^\ast$ enough so that it encompasses two weak links, substantiating once more that only weak links at a distance less than $n^\ast$ contribute to a suppression of the relaxation and they do so by suppressing the rate by $J_{\rm eff}^2$ per weak link. These results are again consistent with the steps observed in the FGR rate (see Fig.~\ref{fig:BoundaryspecWeaklink}), where the number of active weak links scales with the impurity potential.  Finally, the effective weak link couplings can be chosen consistently with the boundary spin $V$, by fixing them to the SW value $J_{\rm eff}=1/2V$. Figure~\ref{fig:birkhofflinksselfconsistent} shows the LIOM decay rate, defined as the plateau value of $\Gamma_{\rm opt}^2$, for different impurity configurations and completes the picture.Similarly to the analysis of the FGR rate, we see that finite density of impurities simply shifts localization/delocalization crossover at given $V$ to somewhat larger system sizes. We emphasize again that this instability is associated not with proliferation of resonances but with factorially growing number of virtual transitions encoded in the operator spreading. 

As for a single impurity, the breakdown of localization can be understood from the flowing localization length. A careful argument put forward in Ref.~\cite{DeRoeck2017avalanche}, known as an avalanche instability, states that if the correlation length of the LIOMs $\xi$ becomes larger than a constant of the order of the lattice spacing the localized phase becomes unstable to unbounded growth of any ergodic seed. One can thus alternatively interpret delocalization of the impurity spins at any disorder strength as an avalanche induced by a flowing localization length $\xi(n)$ with the distance $n$. While we only established the flow of $\xi(n)$ in the weak coupling limit to the bath $\epsilon$, it does not look plausible that the situation changes in the higher orders in $\epsilon$. Indeed for $\epsilon\ll 1$ the shape of the LIOM is $\epsilon$ independent, so one would have to imagine very exotic scenarios where $\xi(n)$ is a non-monotonic function of the distance to stabilize LIOMs. The flow of $\xi(n)$ was observed numerically in two recent works which study the decay rate of the slowest operators in fully disordered models (see Fig. 2 in Ref.~\cite{sels2021markovian} and Fig. 5 in Ref.~\cite{morningstar2021avalanches}). Moreover, a careful analysis of the numerical data in earlier papers claiming to see the exponential scaling of the LIOMs (constant correlation length) reveals that it actually flows considerably with the system size again in agreement with our results (see for example Fig. 2 in Ref.~\cite{Obrien2016Lbits} and Fig. 2 in Ref.~\cite{Pancotti2018Lbits}).

\section{Level statistics and fidelity susceptibility}

In the discussion above we established that the localization-delocalization crossover of a single impurity coupled to a bath is only weakly affected by the presence of other impurities, i.e. by the presence of a disorder potential. Let us now look into two other independent measures, both popular probes to study localization in disordered models. The aim of this section is to establish that these measures agree with our previous analysis, i.e.  that they show same qualitative behavior for a single impurity as for fully disordered models. Because we will study the system as a whole, i.e. without using a probe spin we consider a setup where a single impurity is added in the middle of the chain such that
\be
V_j=V \delta_{j, \ell},\quad \ell=\frac{L+1}{2}
\ee
where $L$ is the chain size which we choose to be odd~\footnote{Some figures will also show even system sizes, in which we set $\ell=L/2+1$}. This problem was studied earlier in the literature~\cite{gubin2012quantum,torres2014localquench,Torres_Herrera_2015, crowley2021partial} focusing in the regime $V\sim 1$. Here we again concentrate
in the regime of large $V$ analyzed above.

\label{sec:SingleImp}
\begin{figure}[htb]
	\centering
	\includegraphics[width= 0.48\textwidth]{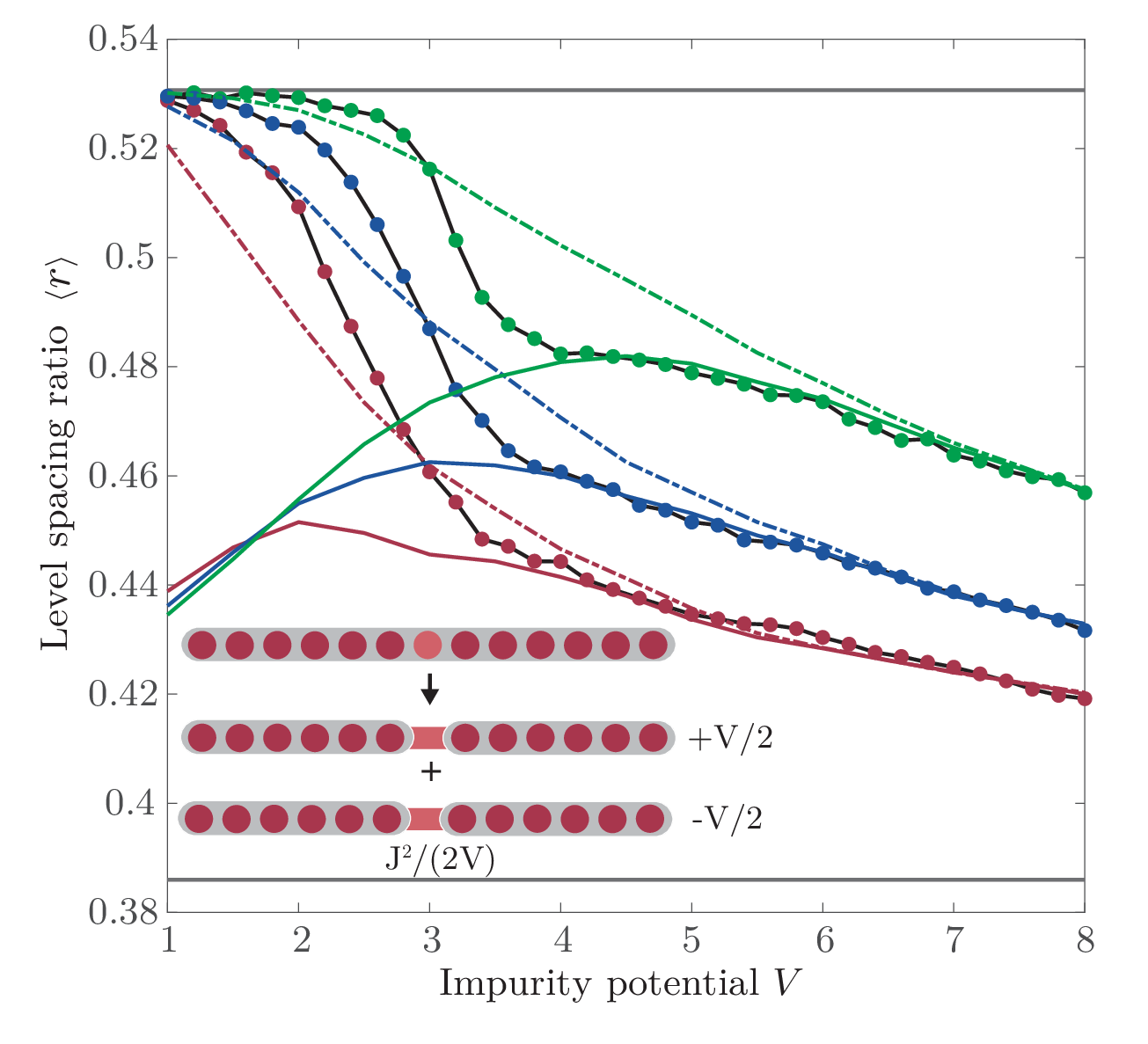}
	\caption{\textbf{Level spacing statistics:} Mean ratio of energy level spacings $\langle r \rangle$ as a function of the impurity potential for Heisenberg chains of length $L=13,15,17$ (red, blue, green) with a single impurity on the central site (black lines with circles). The dashed-dotted lines show the level spacing ratio $\langle r \rangle$ for the effective model where the impurity has been frozen and right-left side of the chain only interact through a virtual process involving the impurity. The full lines result from folding the spectrum of the effective model, resulting from the two possible energies associated with the conserved charge of the impurity.}
\label{fig:levelRsingle}
\end{figure}

\subsection{Level Spacing Statistics}
 
 Fig.~\ref{fig:levelRsingle} shows the mean ratio of energy level statistics as a function of the impurity potential for three different short chains of length $L=13, 15, 17$. Given subsequent energy level spacings $s_n = E_{n+1}-E_{n}$, with $H=\sum_n E_n \left| n\right> \left< n \right|$, this ratio is defined as
\begin{equation}
r_n = \frac{\min(s_n,s_{n+1})}{\max(s_n,s_{n+1})}.
\end{equation}
For non-ergodic systems and Poissonian level statistics, the average over eigenstates $\langle r \rangle\approx 0.386$, whereas for chaotic systems with GOE statistics  $\langle r \rangle\approx 0.5307$~\cite{atas2013distribution}. At sufficiently small impurity potential $V$, the system is observed to be ergodic, as expected. Upon increasing the potential $V$, ergodicity gets broken in a seemingly two-step way. First, there is a fast drop in $\left<r\right>$, followed by a much slower further decrease of the level spacing ratio to the Poissonian value. Furthermore, the required $V$ for the initial deviation from the GOE value shifts significantly with system size $L$. This initial drop is caused precisely by localization of the impurity happening at extensive (up to log corrections) $V^\ast\propto L$ and agrees with the FGR and Birkhoff predictions for the localization threshold. The further slow decay of $\left<r\right>$ is a consequence of the resulting fragmentation of the chain, which occurs at much larger potential $V^{\ast\ast}\propto 2^{L/2}$. So there is a parametrically large window $V^\ast\ll V\ll V^{\ast\ast}$ where the impurity is localized and yet the rest of the system is ergodic. Thus the single impurity model is a specific example of a system with a parametrically large difference between the potentials required to localize the impurity spin and to fragment the Fock space into several (three for our setup) disconnected sectors~\cite{De_Tomasi_2019, detomasi2020rare}. 

To understand the emergence of the asymptotic behavior of $\langle r\rangle$ at large $V$ it is convenient to analyze the effective spin model~\eqref{eq:Heff}, where the impurity spin is integrated out via a Schrieffer-Wolff transformation.  The level spacing statistics of the effective model is illustrated by the dash-dotted lines in Fig.~\ref{fig:levelRsingle}. At sufficiently large impurity potential $V$ they asymptote the full model. Note that at large $V$ the full model is better approximated by an unfolded effective Hamiltonian $H'+V S_\ell^z$, which consists of two decoupled identical blocks, corresponding to the different values of the conserved magnetization of the impurity spin. The level statistics of this unfolded Hamiltonian is illustrated in Fig.~\ref{fig:levelRsingle} by full lines. At very large values of $V$ the separate blocks, corresponding to different values of $S^z_\ell$, do not overlap and the level statistics of the folded and unfolded Hamiltonians are the same, such that dashed and solid lines asymptotically approach each other. As $V$ decreases the impurity still remains frozen such that the effective model is still accurate but the two blocks start to overlap pushing the level statistics closer to the Poisson value. And indeed we see that the solid lines much better approximate $\langle r\rangle$ of the full model. As $V$ decreases further the impurity gets delocalized in the full model such that $\langle r\rangle$ approaches the GOE ratio, and so does the effective model. However, the unfolded Hamiltonian always consist of two decoupled blocks and as they overlap more and more with decreasing $V$, $\langle r\rangle$ is pushed down closer and closer to the Poisson value. We thus conclude that the domain of agreement between the data coming from the full model and the unfolded effective model corresponds to the localized impurity regime. The initial drop in $\left<r\right>$ in the full model from the GOE value is therefore associated with localization of the impurity. The remaining physics can be understood within the effective model. The fact that the magnitude of the jump in $\langle r\rangle$ decreases with the system size is consistent with the expectation that $V^\ast$ corresponding to the freezing of the impurity scales approximately linearly with $L$. In this case the energies of two blocks are extensively separated leading to a very small overlap between the corresponding energies and hence a small drop of $\langle r\rangle$.

\begin{figure}[htb]
	\centering
	\includegraphics[width= 0.48\textwidth]{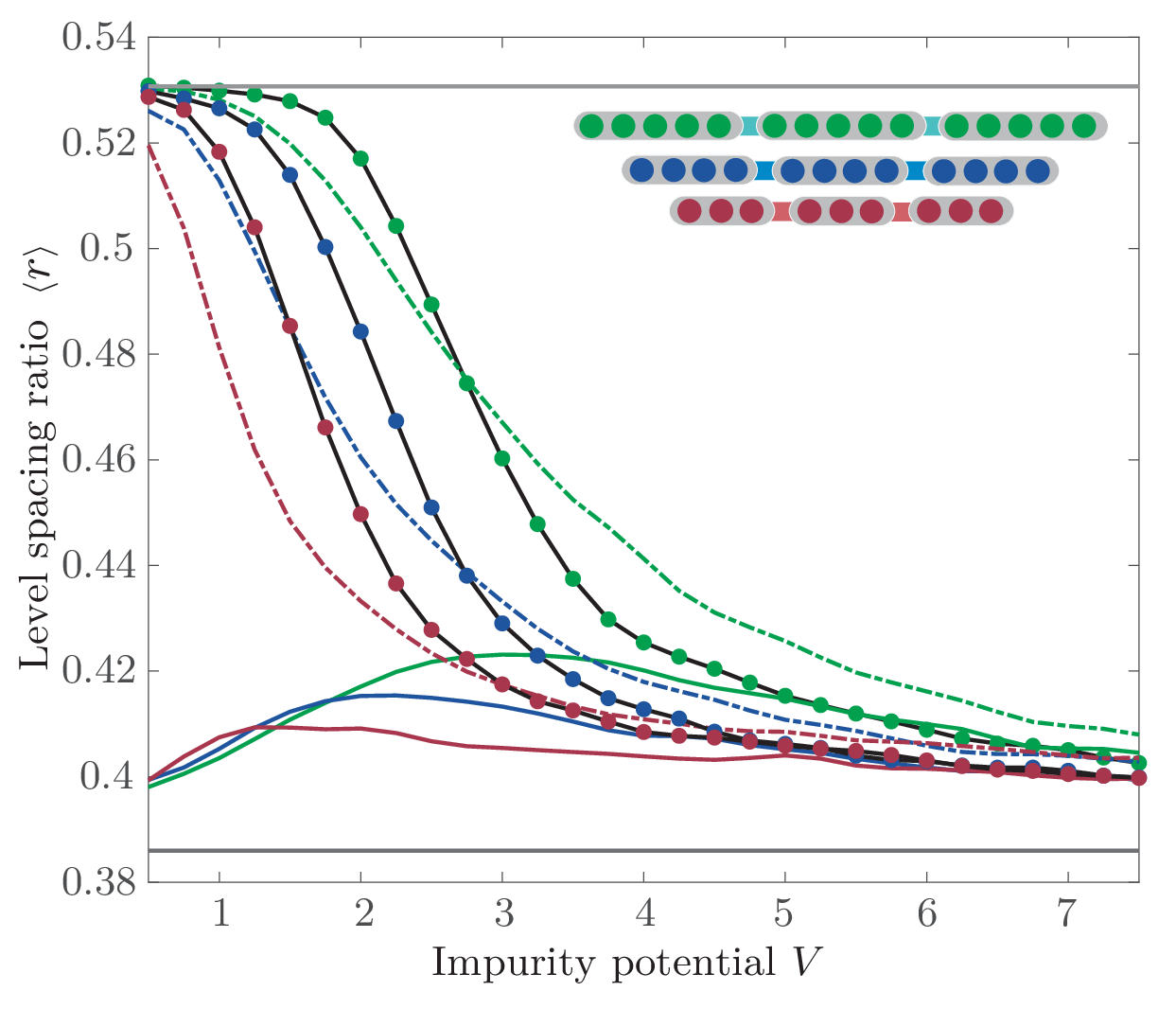}
	\caption{\textbf{Level spacing two impurities:} Mean ratio of energy level spacings $\langle r \rangle$ as a function of the impurity potential for Heisenberg chains of length $L=11,14,17$ (red, blue, green) with a two impurity equally dispersed through the chain (black lines with circles). The dashed-dotted lines show the level spacing ratio $\langle r \rangle$ for the effective model where the impurities have been frozen. The full lines result from folding the spectrum of the effective model, resulting from the four possible energies associated with the conserved charge of the impurities.}
\label{fig:doubleR}
\end{figure}

To highlight the significance of finite size effects on the interpretation of numerical results, we briefly analyze a two-impurity configuration.  Figure~\ref{fig:doubleR} shows the level spacing ratio for the model with two impurities of opposite strength $V$ and $-V$ located as shown in the inset. Comparing these results to Fig.~\ref{fig:levelRsingle}, it becomes immediately clear that the drift in the impurity freezing remains similar, however the drop in level repulsion becomes significantly larger. The latter is easy to understand, as there are now four decoupled blocks once the two impurities have localized. Moreover, because two impurities have been frozen out, the remaining effective model becomes non-ergodic at a smaller impurity potential $V$, which has the same exponential scaling with the system size, but with considerably enhanced finite size effects. For two smaller system sizes (red and blue) one even observes a crossing at large values of $V$, which is often interpreted as a signature of the localization transition~\cite{abanin2019review}. Spectral folding artificially pushes statistics of levels closer to the Poisson value due to overlapping blocks (solid lines) and thus additionally increases finite size effects. Contrary the effective model with frozen impurities shows no signatures of the level crossing and a clear drift of $\langle r\rangle$ towards the GOE value with increasing $L$.

\begin{figure}[ht]
	\centering
	\includegraphics[width= 0.48\textwidth]{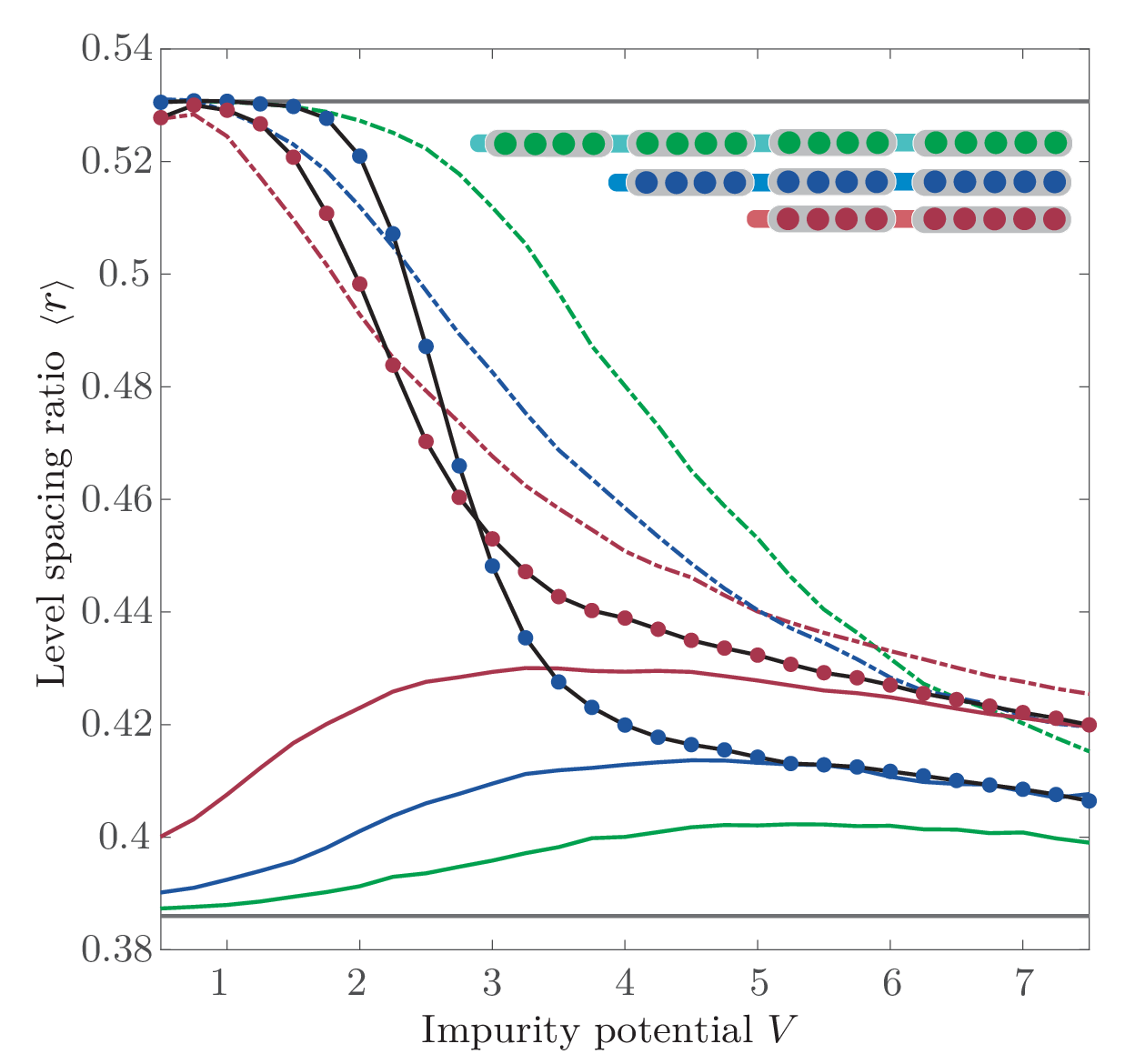}
	\caption{\textbf{Finite density level spacing:} Mean ratio of energy level spacings $\langle r \rangle$ as a function of the impurity potential for Heisenberg chains of length $L=10,15$ (red, blue) with two and three impurities respectively (black lines with circles). The dashed-dotted lines show the level spacing ratio $\langle r \rangle$ for the effective model where the impurity has been frozen. The full lines result from folding the spectrum of the effective model, resulting from the four possible energies associated with the conserved charge of the impurities. For four impurities, with $L_{\rm eff}=17$, the effective model is shown in green.}
\label{fig:Rfixedl}
\end{figure}

We move on to analyzing a multiple impurity setup corresponding to a constant spacing between them $\Delta \ell=5$. The corresponding dependence of $\langle r\rangle$ on $V$ is plotted in Fig.~\ref{fig:Rfixedl}. The black lines and dots show the results for the full models of sizes $L=10$ and $L=15$ as illustrated in the inset. The largest system size corresponding to the green configuration has $L=20$ and is outside of reach of exact diagonalization. Nevertheless we can extrapolate the other two lines noting the drift to the right of the departure from the GOE statistics on top and drift to the left of the departure from the folded effective model (full colored lines). This extrapolation would almost certainly lead to a good crossing point with the two other sizes at $V\approx 2.9$. However, we see that this feature is an entirely spurious effect. It comes from the real drift of the drop position in statistics to larger values of $V$ with $L$ and simultaneous increase in the drop magnitude with $L$ coming from increasing number of effective blocks corresponding to different frozen impurity arrangements. If we look into the effective model folded or unfolded we see a very clear indication that the effective model is ergodic with no crossing in $\langle r\rangle$ developing in the unfolded model and a crossing strongly drifting to the larger values of $V$ with $L$. 

\subsection{Fidelity Susceptibility}

\begin{figure}[htb]
	\centering
	\includegraphics[width= 0.48\textwidth]{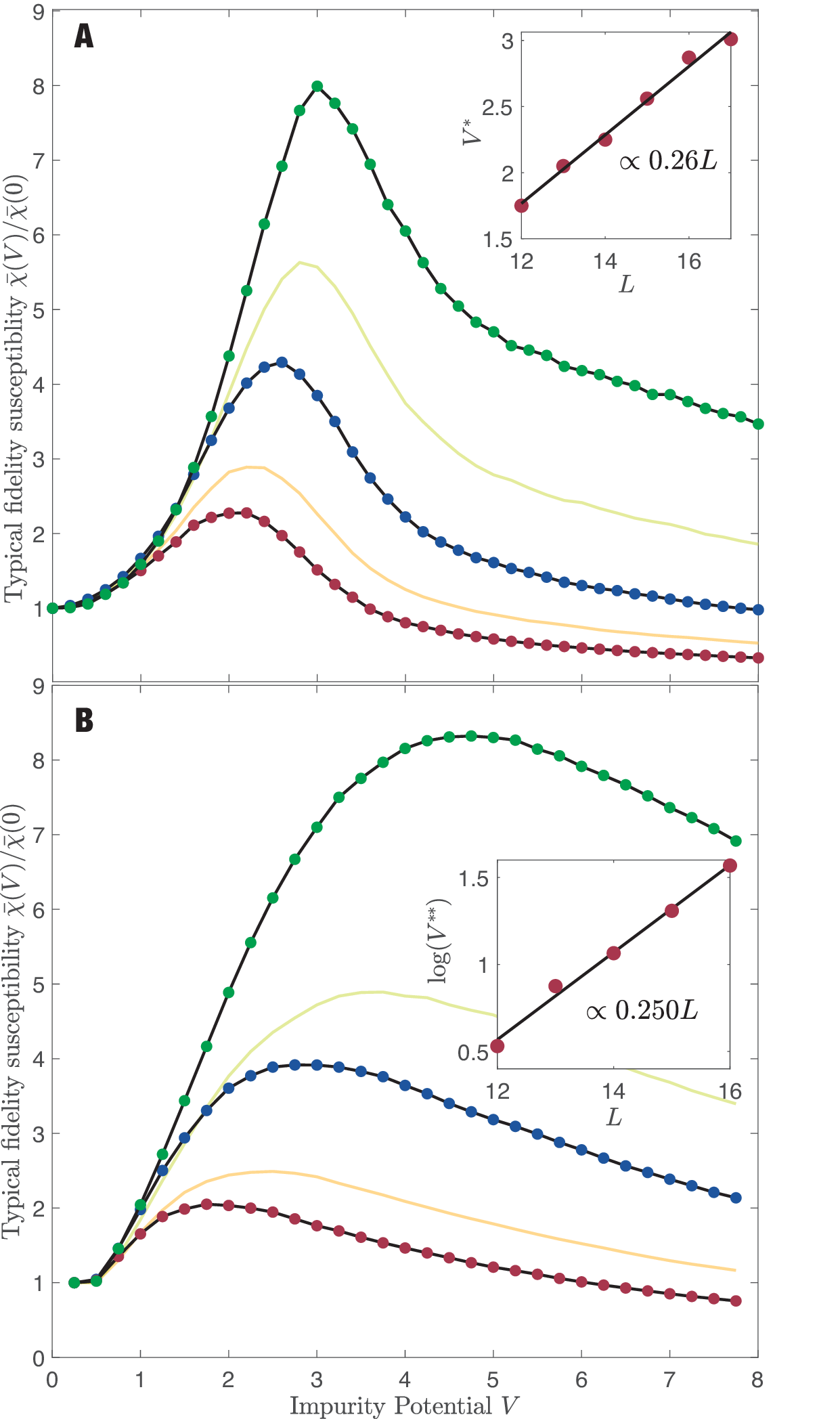}
	\caption{\textbf{Typical Fidelity susceptibility:} Panels A and B show the typical fidelity susceptibility scaled by its value in the absence of an impurity, i.e. at $V=0$. Different system sizes $L=13,15,17$ are shown in red,blue,green respectively. In addition we show $L=14,16$ in yellow and green. Panel (A) corresponds to the susceptibility of a bulk spin in the full model, whereas  panel (B) shows the susceptibility of the same spin in the effective model. The insets in panels (A)/(B) show the scaling of the peak position with system size, together with the best linear/exponential fit. Physical system sizes corresponding to the full and effective models shown in the same color are identical, but the as the impurity spin in the effective model is frozen its actual system size is reduced by one.}
\label{fig:typicalAGP}
\end{figure}

The fidelity susceptibility $\chi$, or equivalently the diagonal component of the quantum geometric tensor with respect to some coupling $\lambda$ can serve as a very sensitive probe of quantum chaos~\cite{pandey2020chaos,leblond2020universality,sels2020dynamical}. Specifically, it has been established that at the crossover from an integrable to an ergodic regime the fidelity susceptibility saturates its upper bound, diverging with the system size as $\chi\propto \exp[2 S(L)]$, where $S(L)$ is the infinite-temperature entropy of the system. For comparison, in integrable regimes $\chi$ diverges at most polynomially with the system size and in the ergodic regime it diverges as $\exp[S(L)]$. For a given eigenstate $n$ the fidelity susceptibility is defined as~\cite{venuti2007geometrictensor, kolodrubetz2017geometry}
\be
\label{eq:chi_n}
\chi_n=\langle n|\overleftarrow{\partial_\lambda} \partial_\lambda|n\rangle_c\equiv  \sum_{m\neq n}{|\langle n |\partial_\lambda H |m\rangle|^2\over (E_n-E_m)^2}.
\ee
For concreteness, we use the longitudinal magnetization of the spins in the bulk of the system as a probe, i.e. $\partial_\lambda H=S^z_3$. To avoid dealing with large fluctuations due to the broad distribution of $\chi_n$ in the non-ergodic phase, we look at the typical susceptibility, defined as 
\begin{equation}
    \chi=\exp(\mathbb{E}[\log \chi_n]),
\end{equation}
where the expectation is over all eigenstates and realizations of the weak disorder in the chain. It is convenient to scale $\chi$ by the ergodic value corresponding to $V=0$ and analyze the ratio $\chi(V)/\chi(0)$, which should saturate at $L$ independent value in the ergodic regimes and diverge exponentially at the localization transition. This scaled susceptibility is plotted in Fig.~\ref{fig:typicalAGP}.

In Fig.~\ref{fig:typicalAGP} (A) we illustrate the susceptibility for the full model. At small values of $V$, i.e. on the ETH side, we identify a good collapse of the data followed by a clear peak in the susceptibility with a height that approximately scales like $\chi\sim e^{2S(L)}$. The inset shows the extracted peak position with system size, the latter is linear to good approximation with a numerically extracted slope $V^\ast \propto 0.26 L$. This expectation up to a $\log(L)$ correction fully agrees with the scaling extracted earlier comparing the FGR relaxation rate and the level spacing (see Eq.~\eqref{eq:V_FGR}). The $\log(L)$ correction is not visible in numerics due to small system sizes. Further note that the level spacing ratio $\left< r\right>$ (see Fig.~\ref{fig:levelRsingle}) at the peak susceptibility is close to the GOE value. The latter is consistent with recent works on MBL~\cite{suntajs2019quantum,suntajs2020transition,sels2020dynamical,leblond2020universality}.

In Fig.~\ref{fig:typicalAGP}~(B) we perform the same analysis on the effective model~$H'$. Note that folding does not affect $\chi$ as the eigenstates in both blocks do not talk to each other. Once more, we observe a peak in the susceptibility, indicating ergodicity breaking in the effective model. However, this time the peak develops much slower and as such appears to drift much faster with the system size. Again, for available system sizes the peak happens at a rather high value of $\left<r\right>$, where there is still a considerable difference in $\langle r\rangle$ between the folded and unfolded models. The inset shows the drift of the peak position on a log-scale with the best fit. This drift is well approximated with a linear curve, indicating this time that the critical interaction needed to decouple the effective model into the independent left and right blocks scales exponentially with $L$. The standard expectation, following from many-body perturbation theory, is that the strength of the effective hopping $J^{\rm eff}=1/(2V)$ coupling two blocks of length $L/2$ sufficient for thermalization scales as $J^{\rm eff}\sim\exp[-S(L)/2]=2^{-L/2}$~\cite{d2016quantum, leblond2020universality, crowley2021partial}. Mathematically, this criterion comes from requiring convergence of the leading perturbative correction to eigenstates and an assumption that the spectral function of the perturbation $\partial_\lambda H$ is flat at small frequencies. The latter assumption is indeed correct (see the inset in Fig.~\ref{fig:Xboundary}). This criterion would predict that $V^\ast\propto \exp[L\log(2)/2]\approx \exp[0.35 L]$, which gives a somewhat larger slope than that in the inset of Fig.~\ref{fig:typicalAGP} (B). The discrepancy could be due to small system sizes leading to the small dynamical range and/or relevance of various $\log(L)$ corrections affecting the observed scaling.

\begin{figure}[ht]
	\centering
	\includegraphics[width= 0.48\textwidth]{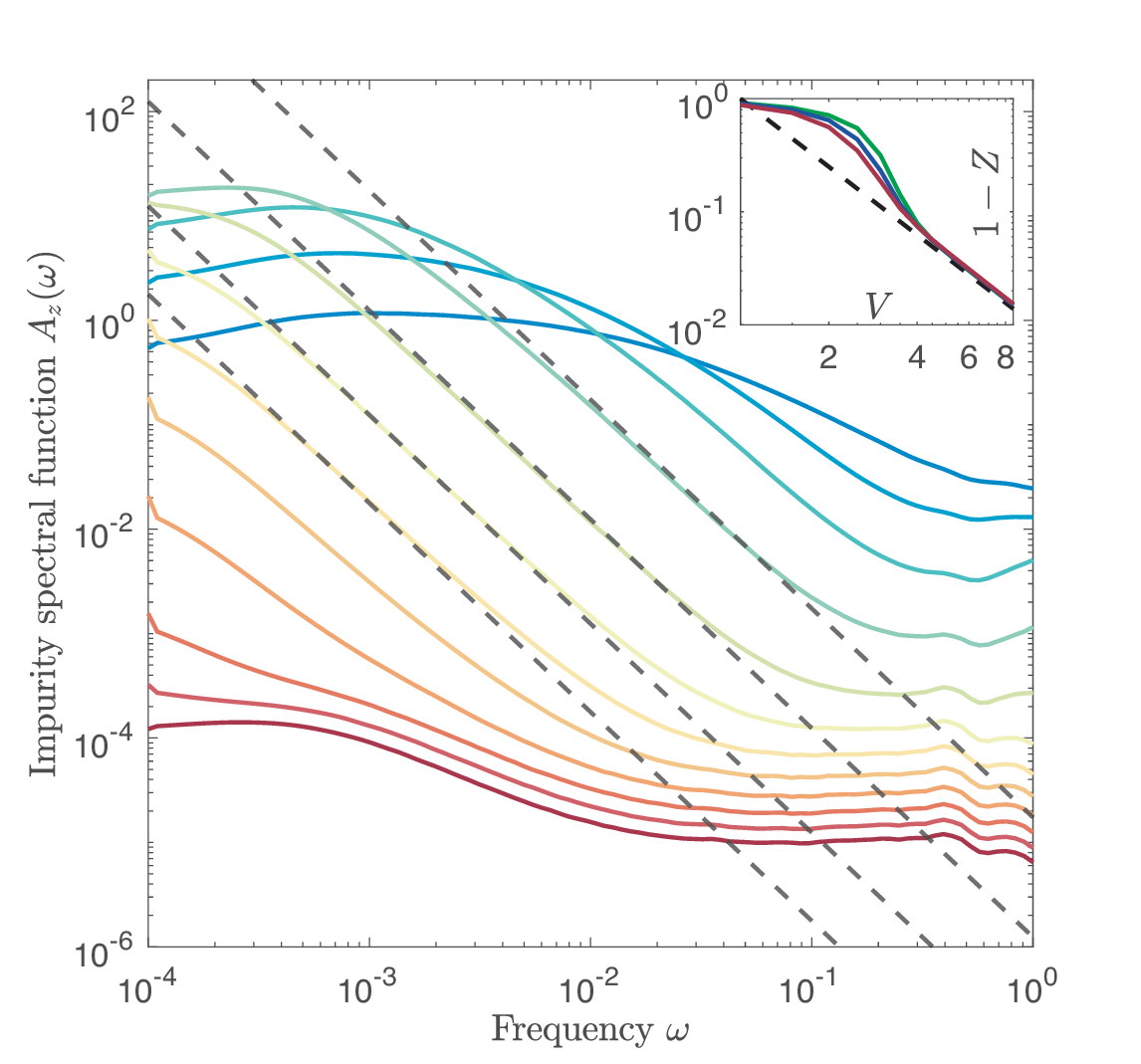}
	\caption{\textbf{Impurity spectrum:} For impurity potentials ranging from $V=1$ to $V=8$, the spectral function of the impurity is shown from blue to red in a system of $L=17$ spins. Dashed lines are guides for the eye and indicate $1/\omega^2$ scaling. The inset shows $1-Z$, where $Z=4 \mathbb{E}[\left<n|S^z_0|n\right>^2]$ with the expectation over all eigenstates and realizations of the weak disorder.}
\label{fig:specZimp}
\end{figure}

Exponential enhancement of the fidelity susceptibility implies an exponential (in $L$) enhancement of spectral weight at low frequency from $O(1)$ to $O(\exp(S(L)))$, accompanied by exponentially slow (in $L$) relaxation~\cite{pandey2020chaos, sels2020dynamical}. To confirm that this is the case it is thus instructive to look directly at the spectral function of the impurity, which is shown in Fig.~\ref{fig:specZimp} for various strengths of the impurity potential $V$. At intermediate $V$, before a significant fraction of the magnetization has become conserved and the associated amount of spectral weight has been transferred to $\omega=0$, we observe a clear $1/\omega^2$ scaling at low frequencies. The latter was recently observed in other systems with slightly broken integrability~\cite{schonle2020eigenstate, leblond2020universality}. This scaling is indicative of Lorentzian line broadening.
In turn, the Lorentzian shape of the spectral functions suggests that the relaxation of $S_\ell^z$ is simply governed by Fermi's golden rule (FGR) (see the last Appendix in Ref.~\cite{leblond2020universality} for a detailed discussion). In passing we note that the spectral function of the bulk spin defining the fidelity susceptibility plotted in Fig.~\ref{fig:typicalAGP} shows slower $1/\omega$ subdiffusive scaling behavior. It is illustrated in Fig.~\ref{fig:bulkspecapp} in Appendix F and agrees with the results reported by us earlier in Ref.~\cite{sels2020dynamical} for a fully disordered model. This $1/\omega$ scaling corresponds to a very slow, logarithmic in time, relaxation which is somewhat surprising for the effective model.

\section{Conclusion and outlook}
In this work we have presented a numerical and analytical study of one dimensional Heisenberg chains with dilute sets of defects, being spins with a large external field. We first analyzed the crossover from a localized to a delocalized regime for a single probe impurity weakly coupled to an ergodic bath. We showed that this crossover can be explained from the ergodic side by comparing the FGR decay rate and the mean level spacing of the bath and from the localized side by the divergence of the Birkhoff construction of the LIOM connected to the impurity spin. Interestingly, both approaches give the same criterion for the localization/delocalization crossover.

We tied the divergence of the Birkhoff construction to the Krylov complexity of the bath. In local interacting models (disordered or not) this complexity saturates its upper bound resulting in (almost) factorial growth of norms of nested commutators and as a result to the instability of the LIOM. In this way we avoid any need of making any assumptions about the eigenstates of the bath and can work directly in the thermodynamic limit. Thus we concluded that adding a finite density of disordered sites does not affect the fact that in the thermodyanmic limit there is no localized phase, but does quantitatively affect both the time scales at which impurity delocalizes and the length scale of the crossover between the localized and delocalized regimes. Let us comment that MBL is often argued to be related to localization on graphs like random regular graphs~\cite{Kravtsov_RRG_2015} or Caley trees~\cite{Kravtsov_Bethe_2018}. From the point of view of the Birkhoff construction there is a huge qualitative difference between them and the local models. The nested commutator norms $\|R_k\|$ on such graphs can only grow exponentially with $k$ such that the Birkhoff construction converges at a sufficiently large impurity potential (see Eq.~\eqref{eq:Gamma_N}) even if the bath is not disordered, i.e. ergodic. So a weakly coupled impurity to such a system at a sufficiently large $V$ would be localized (at least in the small $\epsilon$ limit). Adding disorder to the system will simply shift the localization transition to a smaller value of $V$.

We also analyzed numerically various other proxies for egodicity, such as level spacing ratio's, fidelity susceptibilities and spectral functions. All these measures point to the same conclusion that, regardless of the impurity density or the potential, the impurities ultimately relax in the thermodynamic limit by dissipating energy in the remaining bath. Nonetheless the dynamics is exponentially slow in $V$. 

Our conclusions are opposite to previous works which argue for the stability of the MBL phase based on the analysis of the effect of resonances~\cite{basko2006MBL, gornyi2005MBL, imbrie2016MBL,Pietracaprina2016}. The physical mechanism of instability, which we found here, is based on virtual non-resonant processes and is ultimately tied to the operator growth which is absent in non-interacting systems. This instability develops at $V\propto L/\log(L)$, which corresponds to energies, where resonances cannot play a role simply because of a small density of states near the edge of the many-body spectrum. Of course, perturbative divergence of the decay rate does not exclude that there are some other, non-perturbative mechanisms stabilizing LIOMs. But given that our analytical predictions fully agree with all known to us numerical data, as well as with the variational approach (see Appendix C), we find this scenario very unlikely. 

We believe that our analysis is fully consistent with most, if not all, numerical results on the MBL transition. In particular, it explains (i) the approximately linear drift of transition when identified close to the GOE limit, as recently suggested by~\cite{suntajs2020transition,Assa2016,sels2020dynamical}, (ii) the crossing point in level spacing statistics with apparent very slow drift~\cite{Pal2010MBL,De_Tomasi_2019}, (iii) the non-monotonic behavior of ergodicity probes with system size~\cite{Sierant_2020}, (iv) the low energy tail developing in the spectral function in the localized regime~\cite{Serbyn2017Thouless,sels2020dynamical} manifested in slow subdiffusive transport~\cite{Zniraric2016Subdiffusion,Kratiek2015,luitz2017subdiffusion,Doggen2021ManybodyLI}. 

Our work suggests that in thermodynamic limit instead of MBL there is a transient glassy-type phase characterized by finite-time subdiffusive, logarithmic in time, spreading of correlation functions. Such slow transport was previously attributed to the MBL phase~\cite{Bardarson_MBL_2012}. Heuristically one can think about this transient regime as a stage of slowly dephasing ``quasi-particles'' or l-bits/LIOMs, which eventually crossovers to their subsequent diffusion. Such a crossover from sub-diffusive to faster transport is not unique to disordered systems and was observed in other setups, see e.g. Refs.~\cite{Zvonarev_Spin_2007,Ponno_FPU_two_stage_2011,howell2019prethermalization}

\acknowledgements \emph{Acknowledgements.} The authors would like to thank A. Chandran, P. Crowley and D. Huse, T. Prosen, M. Rigol, L. Vidmar for useful discussions related this work. The Flatiron Institute is a division of the Simons Foundation. D.S. was supported by AFOSR: Grant FA9550-21-1-0236. A.P. was supported  by NSF: Grants DMR- 1813499 and DMR-2103658 and by AFOSR: Grants  FA9550-16-1-0334 and FA9550-21-1-0342

\bibliography{ref_general}

\section*{Appendix A: Connection between $\chi$ and $g=\Gamma/\Delta$.}
 Within the context of many-body localization, the dimensionless coupling $g=\Gamma/\Delta$, being the ratio of the Fermi golden rule rate $\Gamma$ and the level spacing $\Delta$, has been proposed as a measure for ergodicity~\cite{vosk2015RG}. In single particle systems this ratio defines the the dimensionless Thouless conductance. Ergodic systems are usually characterized by a $\log(g)\sim cL$, where $c>0$ and $L$ is the system size. The localized, non-ergodic, phase is characterized by $g \rightarrow 0$ in the thermodynamic limit, typically one would expect $\log(g) \sim cL$ where $c<0$. Colloquially speaking $g$ measures whether or not there are sufficiently many states within the line width for Fermi's golden rule to hold.
 
 The purpose of this appendix is simply to point out that the fidelity susceptiblity $\chi_n$ of an eigenstate $\left| n\right>$ as defined by~\eqref{eq:chi_n}, under some reasonable assumptions, is equivalent to the dimensionless coupling $g_n$ of that eigenstate. In the present context it's most useful to consider the FGR rate, and susceptibility, for connecting spatially disconnected blocks together but before we do so we will present some general result.
 
 Consider expression~\eqref{eq:chi_n}, it can be rewritten as 
 \begin{equation}
     \chi_n(\lambda)=\int {\rm d}\omega \frac{A_n(\omega;\lambda)}{\omega^2},
     \label{eq:chi_appendix}
 \end{equation}
 where the spectral function is defined as 
 \begin{equation}
     A_n(\omega;\lambda)=\sum_{m\neq n} |\left< n | \partial_\lambda H |m\right>|^2 \delta(\omega-(E_n-E_m)).
 \end{equation}
As long as the spectral function tends to a constant at low frequency, the integral~\eqref{eq:chi_appendix} is infra-red divergent and completely dominated by the small denominators. As such, the typical susceptiblity becomes
\begin{equation}
    \chi_n \approx \frac{A_n(0^+)+A_n(0^-)}{\Delta},
\end{equation}
where $\Delta$ is the typical level spacing. On the other hand, the numerator is  directly related to the FGR decay rate of the eigenstate $\left|n\right>$ upon perturbing it with $\partial_\lambda H$ , i.e. 
\begin{equation}
    \Gamma_n=2\pi(A_n(0^+)+A_n(0^-)).
\end{equation}
Consequently, we arrive at the rather straightforward conclusion that 
\begin{equation}
\label{eq:chi_Gamma}
    \chi_n \approx 2\pi \frac{\Gamma_n}{\Delta}= 2\pi g_n.
\end{equation}
The equivalence is thus expected to hold as long as the system is ergodic, where the spectral function has a robust low frequency plateau. Let us emphasize that the susceptibility $\chi_n$ entering Eq.~\eqref{eq:chi_Gamma} is computed in the limit of an infinitesimal coupling of the impurity to the bath. As coupling increases the spectral function gets strongly renormalized quickly reaching the maximum value $\chi_n\sim 1/\Delta^2$~\cite{leblond2020universality} and then decreasing back to the expected ETH scaling $\chi_n\sim 1/\Delta$.

To be specific, let's consider the Hamiltonian 
\begin{equation}
    H=H_L+V S^z_{\ell}+ \lambda S^x_{\ell-1} S^x_\ell, 
\end{equation}
where $S^x_{\ell-1}$ is the boundary spin of the Hamiltonian $H_L$. The latter is coupled to a spin $\ell$ with external field $V$. In the decoupled limit, when $\lambda=0$, the spectral function for coupling $\partial_\lambda H$  between the system $H_L$ and the new spin becomes
\begin{equation}
    A_n(\omega;0)=\int {\rm d} \nu X^{(\ell-1)}_n(\omega-\nu)  X_n^{(\ell)}(\nu),
\end{equation}
where $X^{(\ell-1)}_n$ denotes the spectral function of the $S^x_{\ell-1}$ and $X^{(\ell)}_n$ denotes the spectral function of the newly coupled spin. Given that the newly connected spin simply rotates around the z-axis at frequency $V$, we have $X^{(\ell)}_n=(\delta(\omega \pm V)/4$, depending on whether the $\ell^{\rm th}$ spin is up or down. As such we find
\begin{equation}
     A_n(\omega;0)=\frac{1}{4} X^{(\ell-1)}_n(\omega \pm V).
\end{equation}
It follows that the FGR rate for the decay of an eigenstate is simply 
\begin{equation}
    \Gamma_n=\frac{ \pi}{2} X^{(\ell-1)}_n(\pm V).
\end{equation}


\section*{Appendix B: Birkhoff construction of the LIOM}
In this appendix we lay out the recursive (Birkhoff) construction of the LIOM formed out of deformations of the impurity spin. We consider the Hamiltonian~\eqref{eq:H_bi}, where $V$ is large and $\epsilon$ is small. Our goal is to construct a conserved charge in the leading order in $\epsilon$ but in all orders in $1/V$. Any conserved charge should satisfy $[Q,H_{\rm bi}]=0$. Consider some iterative scheme where one has an estimate $Q_n$ of the conserved charge in the $n$-th iteration with $Q_0=S^z_0$.  This charge won't exactly be conserved, let's say there is some residual operator
\begin{equation}
T_n=[Q_n,H].
\end{equation}
Now we can ask whether there is an operator, which we could add to $Q_n$ such that it would cancel the residual $T_n$ when commuted with $VS^z_0$, i.e.
\begin{equation}
[q_n,VS^z_0]=-T_n, 
\end{equation}
such that the new conserved charge becomes $Q_{n+1}=Q_n+q_n$. In the leading order of expansion this scheme gives
\[
[q_1, V S^z_0]=-T_1=-\epsilon [S^z_0, H_{\rm int}],
\]
where we used that $[S^z_0, H_{\rm bulk}]=0$ which yields the solution
\begin{equation}
    q_1={\epsilon\over V} H_{\rm int},\quad Q_2=S^z_0+q_1
\end{equation}
We can now continue this construction. In the next order we need to solve the equation
\begin{equation}
\label{eq:delta2}
[q_2, V S^z_0]=-T_2=-[q_1, H_{\perp}]= -[q_1, H_{\rm bulk}],
\end{equation}
Let us point out that the equation
\[
[X,S^z_0]=A
\]
only admits a solution for $X$ if the operator $A$ is odd under parity transformation generated by $\sigma ^z_0=2S^z_0$: $\sigma^z_0 A \sigma^z_0=-A$. This follows e.g. by multiplying both sides to the equation above by $\sigma^z_0$ on the left and on the right. If this condition is satisfied then it is easy to check that
\begin{equation}
\label{eq:XAB}
    X=-\sigma^z_0 A+B,
\end{equation}
where $B$ is an arbitrary operator commuting with $\sigma^z_0$. Because $H_{\rm int}$ is an even operator and $H_{\rm int}$ is odd, the parity of any nested commutator of these two operators is determined by whether $H_{\rm int}$ appears even or odd number of times. The RHS of Eq.~\eqref{eq:delta2} is obviously odd such that
\begin{multline}
    q_2=-{1\over V}\sigma^z_0 [H_{\rm int}, q_1]=\\
    -{\epsilon\over V^2}\sigma^z_0 [H_{\rm bulk}, H_{\rm int}]=-{\epsilon\over V^2}[S_0^z,[H_{\rm bulk},H_{\rm int}]]
\end{multline}
Here we set the arbitrary commuting operator $B$ to zero, which as it will become clear shortly is justified in the linear order in $\epsilon$. In general $B$ should be chosen to cancel all even terms appearing in $T_n$.

We can now continue this construction iteratively solving the equation 
\begin{equation}
\label{eq:delta_n}
    [q_n,VS^z_0]=-[q_{n-1},H_{\perp}]\approx -[q_{n-1}, H_{\rm bulk}],
\end{equation}
where we replaced $H_{\perp}$ in the RHS of this equation by $H_{\rm bulk}$ because keeping $H_{\rm int}$ would result in $O(\epsilon^2)$ corrections to $q_n$. Using Eq.~\eqref{eq:XAB} it is now straightforward to check that the solution of Eq.~\eqref{eq:delta_n} reads
\begin{equation}
    q_n=-{1\over V}\sigma^z_0 [H_{\rm bulk},q_{n-1}]={\epsilon\over V^n}(-\sigma^z_0)^{n-1} {\rm Ad}^{n-1}_{\rm H_{\rm bulk}} H_{\rm int}.
    \label{eq:delta_n}
\end{equation}
This yields the expansion:
\begin{equation}
    Q_{2n}=S^z_0+{\epsilon\over V}\sum_{k=0}^{2n} \left({-\sigma^z_0\over V}\right)^k {\rm Ad}^k_{H_{\rm bulk}} H_{\rm int},
\end{equation}
which as it is easy to see is equivalent to the expansion~\eqref{eq:expansion_tau_z} in the main text if we relabel $2n\to n$.

Alternatively, one could consider a finite system of size $L$ such that the sum in Eq.~\eqref{eq:expansion_tau_z} converges at sufficiently large $V$ and sufficiently small $\epsilon$. Then, the expression above can be resummed to the infinite order, leading to
\begin{equation}
    Q=S^z_0+{\epsilon\over V}\left({1\over 1+{2 S^z_0\over V}Ad_{H_{\rm bulk}}}\right)H_{\rm int}
\end{equation}
The norm of this operator can be straightforwardly computed in the eigenbasis of the uncoupled Hamiltonian $H_{\rm bulk}+VS^z_0$:
\begin{multline}
    \|Q\|^2={1\over 4}+{\epsilon^2\over 2^L}\sum_{n,m} {\left|\langle n|H_{\rm int}|m\rangle\right|^2 \over (E_n-E_m\pm V)^2} =\\ {1\over 4}+\epsilon^2 \chi,
\end{multline}
where $\chi$ is the eigenstate-average fidelity susceptibility, which we introduced in the previous appendix;
$\pm$ sign refers to ``up'' and ``down'' sectors of the spin $S^z_0$. This result once again leads to the conclusion that the norm of the conserved charge is related to the fidelity susceptiblity, which as it was already shown in the previous appendix is related to the ratio of the FGR to the level spacing. Because the conserved part of magnetization scales as ${\rm Tr (S^z_0 Q)}/{\rm Tr} (Q^2)\sim 1/\|Q\|^2$ we conclude that the condition $\epsilon^2\chi\gg 1\;\leftrightarrow\; \Gamma\gg \Delta$ implies that this conserved magnetization is small.

While in this paper we focused on quantum systems, let us point that this LIOM construction applies to the classical setup, where the Hamiltonian~\eqref{eq:H_bi} is expressed not in terms of spin -1/2 operators but in terms of continuous angular momenta satisfying Poisson bracket relations:
\[
\{S^x_i,S^y_j\}=S^z_i \delta_{ij}
\]
plus cyclic permutations. Then it is easy to check that the Birkhoff construction for the conserved charge $Q$, satisfying $\{Q,H_{\rm bi}\}$=0 in the linear order in $\epsilon$ proceeds as follows:
\begin{equation}
Q_{2n}=S_0^z+\epsilon \sum_{q=0}^{2n} {1\over V^{q+1}} (f_x^{(q)} S^x_0+f_y^{(q)} S^y_0),
\end{equation}
where $f_x^{(0)}=S^x_1$, $f_y^{(0)}=S^y_1$, and for $q>0$:
\[
f_x^{(q)}=\{H_{\rm bulk},f_y^{(q-1)}\},\quad f_y^{(q)}=-\{H_{\rm bulk},f_x^{(q-1)}\}.
\]
The decay rate of this conserved charge is completely analogous to Eq.~\eqref{eq:Gamma_N}, where $R_k={\rm Ad}^{\,k}_{H_{\rm bulk}} H_{\rm int}$ with the ``Ad'' operator implying the nested Poisson brackets. The norm of the function $R_k$ is defined through a phase space average over orientations of all spins: $\|R_k\|^2=\int \prod d\theta_i d\phi_i R_k^2$, where $\theta_i$ and $\phi_i$ are the spherical angles defining spin orientations. As it was argued already in Ref.~\cite{parker2018universal} the scaling of these norms are expected to have the same factorial scaling for generic local Hamiltonians. 

\section*{Appendix C: Variational construction of the LIOM}

The Birkhoff construction in nested commutators suggests how one can go beyond diverging perturbative expansion by considering the following variational ansatz:
\begin{equation}
\label{eq:Q_n_var_1}
    Q_n^{\rm var}=S_0^z+\epsilon \sum_{k=0}^n \alpha_k {\rm Ad}_{H_{\rm bulk}}^{2k} H_{\rm int}+\epsilon \sum_{k=1}^n \beta_k \sigma_0^z {\rm Ad}_{H_{\rm bulk}}^{2k-1} H_{\rm int}.
\end{equation}
As we mentioned in the main text this ansatz is formally exact (in the linear order in $\epsilon$) for any finite system in the limit $n\to \infty$. In order to simplify the analysis 
we can use the approximation
$\beta_k=-V\alpha_k$, which is exact in the perturbative regime and which shows the same qualitative features as the more complete ansatz discussed in the main text. We will first show how one can work directly with the nested commutators and then discuss what happens if we use orthonormal Krylov basis.

The variational solution can be found by minimizing the norm of commutator of the ansatz conserved charge with the Hamiltonian, $\|[Q_n^{\rm var}, H]\|^2$. 
Using the trace properties of products of nested commutators discussed in the main text it is easy to find that
\begin{multline}
\label{eq:var_norm_gamma}
  \Gamma_{n,\, var}^2\equiv \|[Q_n^{\rm var}, H]\|^2=(1+\alpha_0 V)^2\\
    +\sum_{k,q=0}^n (\alpha_q-V^2\alpha_{q+1})(\alpha_k-V^2\alpha_{k+1}) \|R_{k+q+1}\|^2.
\end{multline}
The perturbative Birkhoff solution $\alpha_0=-1/V,\;\alpha_q=1/V^2 \alpha_{q-1}$ clearly emerges in the limit of large $V$. Note that because $\alpha_{n+1}\equiv 0$ all terms cancel except for the last one, leading to Eq.~\eqref{eq:Gamma_N}. However, as $n$ increases at fixed V the variational solution starts to depart from the perturbative one. 

One can further simplify Eq.~\eqref{eq:var_norm_gamma} by changing the variables from $\alpha_k$ to $\tilde \alpha_k=\alpha_k/V^{2k+1}$. This change results in rescaling $\|R_{k+q+1}\|^2\to \|R_{k+q+1}\|^2/V^{2(k+q+1)}$. Using the asymptotic expression for the nested commutator norm~\eqref{eq:norm_nested_avdoshkin} we see that this change is amounts to setting $V=1$ and renormalizing the parameter $\tau\to V\tau$. The variational solution thus becomes a universal function of $V\tau$ in agreement with the perturbative result (see e.g. Eq.~\eqref{eq:Gamma_N}). The requirement that $\tilde \alpha_{n+1}=0$, is equivalent to $\sum_{k=0}^{n+1} \xi_k=0$, which can be enforced through the Lagrange multiplier $\Lambda$ resulting in the minimization of the following quadratic form:
\be
(1+\xi_0)^2+\sum_{k,q=1}^n \xi_q \xi_k \|R_{k+q+1}\|^2-2\Lambda\sum_{k=0}^{n+1}\xi_k
\ee
The minimization is straightforward, resulting in the following expression for the decay rate of the LIOM:
\[
\Gamma_{n,\,\rm var}^2={1\over 1+\sum_{k,q=1}^{n+1} (\hat R^{-1})_{kq}},
\]
where $\hat R^{-1}$ is the inverse of the Hankel matrix $\hat R$ defined by the matrix elements $\hat R_{k,q}=\| R_{k+q-1} \|^2$, $k,q=1\dots n+1$. 

One can rewrite the same simplified variational ansatz In the Krylov space defined in the main text:
\begin{equation}
    Q_n^{\rm var}=S_0^z+\epsilon \psi_0 O_0+\epsilon[H,P_n]-\epsilon V \sigma_0^z P_n,
\end{equation}
where 
\be
\label{eq:P_n_def}
P_n=\sum_{k=1}^n \alpha_k {\rm Ad}_{H_{\rm bulk}}^{2k-1} H_{\rm int}\equiv \sum_{k=1}^n \psi_{2k} O_{2k-1}.
\ee
Here we introduced a new set of variational parameters $\{\psi_{2k}\}$. Next we compute the commutator $[H_{\rm bi},Q_n]$ in the linear order in $\epsilon$. It is convenient to replace $\epsilon H_{\rm int}$ in the Hamiltonian~\eqref{eq:H_bi} with $\epsilon O_0$, which can be done by a simple rescaling of the parameter $\epsilon$
\begin{widetext}
\begin{multline}
    [H_{\rm bi},Q_n]=\epsilon (\psi_0 V-1) \sigma_0^z O_0 +\epsilon \psi_0 b_1 O_1-\epsilon V^2 P_n
    +\epsilon [H_{\rm bulk},[H_{\rm bulk}, P_n]]=\epsilon (\psi_0 V-1) \sigma_0^z O_0\\
    +\epsilon \sum_{k\geq 1} \left(\psi_{2k} (b_{2k}^2+b_{2k-1}^2-V^2) +\psi_{2k-2} b_{2k-2} b_{2k-1}+\psi_{2k+2} b_{2k+1}b_{2k}\right) O_{2k-1},
\end{multline}
\end{widetext}
where we set $\psi_{2k}=0$ for $k\geq n$. Because the set of operators $O_n$ is orthonormal, the norm of this operator is just the sum of squares of the coefficients in front of the operators $O_k$:
\begin{multline}
    \Gamma_{n,\, var}^2/\epsilon^2=(\psi_0 V-1)^2+\sum_{k\geq 1} \Bigl( \psi_{2k} (b_{2k}^2+b_{2k-1}^2-V^2) \\
    +\psi_{2k-2} b_{2k-2} b_{2k-1}+\psi_{2k+2} b_{2k+1}b_{2k}\Bigr)^2
    \label{eq:var_norm_gamma_1}
\end{multline}
Knowing the Lanczos coefficients $\{b_k\}$ allows one to minimize this quadratic form. As before, it is easy to recover the perturbative results in the limit $V\gg b_{2n}$ with 
\[
\psi_0\approx 1/V,\quad \psi_{2k}\approx {b_{2k-2}b_{2k-1}\over V^2}\psi_{2k-2}.
\]
In generic interacting non-integrable one-dimensional systems $b_k\propto k/\log(k)$~\cite{parker2018universal}, which leads to the same scaling of $\Gamma_{n,\, var}$ as discussed in the main text (see Eq.~\eqref{eq:Gamma_N}). When the potential $V$ becomes comparable to $b_{2n}\sim C\, n/\log(n)$ this scaling breaks down and crosses over to much slower decay of the rate with $n$ as shown in the main text. From~\eqref{eq:var_norm_gamma_1} it also becomes clear why the eigenstates of the quadratic form minimizing the decay rate and shown in Fig.~\ref{fig:birkhoffvar} change their structure from positive highly localized states at small $n$ to oscillating delocalized states at large $n$.

\section*{Appendix D: Birkhof construction for periodic driving}
As we discussed analysing the FGR decay, in the rotating frame the impurity problem maps to the Floquet problem. In the Floquet language, the existence of the LIOM is equivalent to the existence of a local Floquet Hamiltonian. While Floquet driving is not the focus of this work, let us briefly show that the Birkhoff's LIOM construction can be applied directly to Floquet systems, where instead of to an impurity, we couple the system to a photon field such that it is described by the Hamiltonian
\[
H_{\rm bF}=H_{\rm bulk}+\Omega\, a^\dagger a+\epsilon (a^\dagger+a) H_{\rm int},
\]
where $\Omega$ is the photon frequency, which plays the same role as the impurity potential $V$, $a$ and $a^\dagger$ are the photon creation and annihilation operators and $H_{\rm int}$ can be either a boundary spin of the bath, $H_{\rm int}=S^x_1$ or the total transverse magnetization $H_{\rm int}=\sum_j S^x_j$ or something else. In the rotating frame in the limit of a large photon number this Hamiltonian becomes periodically driven with frequency $\Omega$. We can now construct the LIOM coupled to the photon number using the same spirit as before:
\[
Q=a^\dagger a+{1\over \Omega} q_1+{1\over \Omega^2}q_2+\dots,
\]
where $q_j$ are functions of $\epsilon$. It is easy to check that in the linear order in $\epsilon$
\begin{multline}
Q=a^\dagger a+\epsilon (a^\dagger+a) \sum_{q=0}^n {1\over \Omega^{2q+1}}Ad^{2q}_{H_{\rm bulk}} H_{\rm int}\\+\epsilon (a^\dagger-a) \sum_{q=1}^{n} {1\over \Omega^{2q}}Ad^{2q-1} _{H_{\rm bulk}} H_{\rm int}.
\end{multline}
This Birkhoff construction has the structure identical to that for the impurity problem. Therefore we can draw identical conclusions about the asymptotic nature of the LIOM. Note that the breakdown of the LIOM in this case indicates that the photon delocalizes and the system heats up.

In this context, it's worth to note that this result questions recent claims about MBL stabilizing down-converters (which are sometimes referred to as discrete time crystals) in the thermodynamic limit~\cite{Yao_2018, khemani_DTC} . It should be noted though that if the driving field couples to a Hamiltonian which does not have the factorial growth of the nested commutator norms, then at least in the leading order in $\epsilon$ the LIOM converges and there is no heating. This happens, in particular, when one adds a small amplitude drive to the Hamiltonian which consists of a sum of mutually commuting terms. To study heating in those systems one has to extend the Birkhoff construction beyond linear (or possibly any other finite) order in $\epsilon$. We note that in this class of systems it was found numerically that heating is indeed very strongly suppressed~\cite{prosen_98, D_Alessio_2013,Haldar_2018}

\section*{Appendix E: Spectral functions for the full vs. effective models.}

\begin{figure}[ht]
	\centering
	\includegraphics[width= 0.45\textwidth]{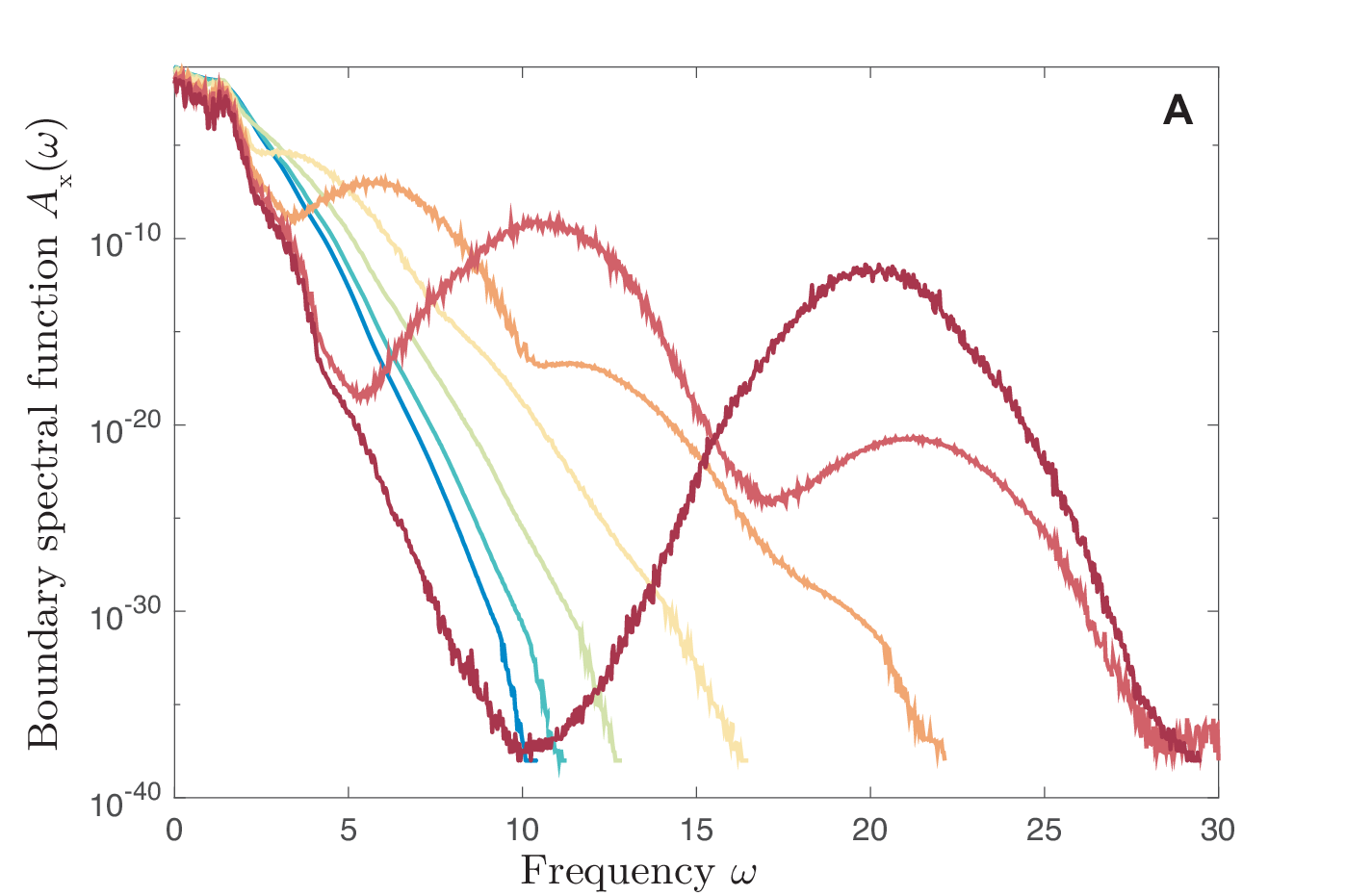}
	\includegraphics[width= 0.45\textwidth]{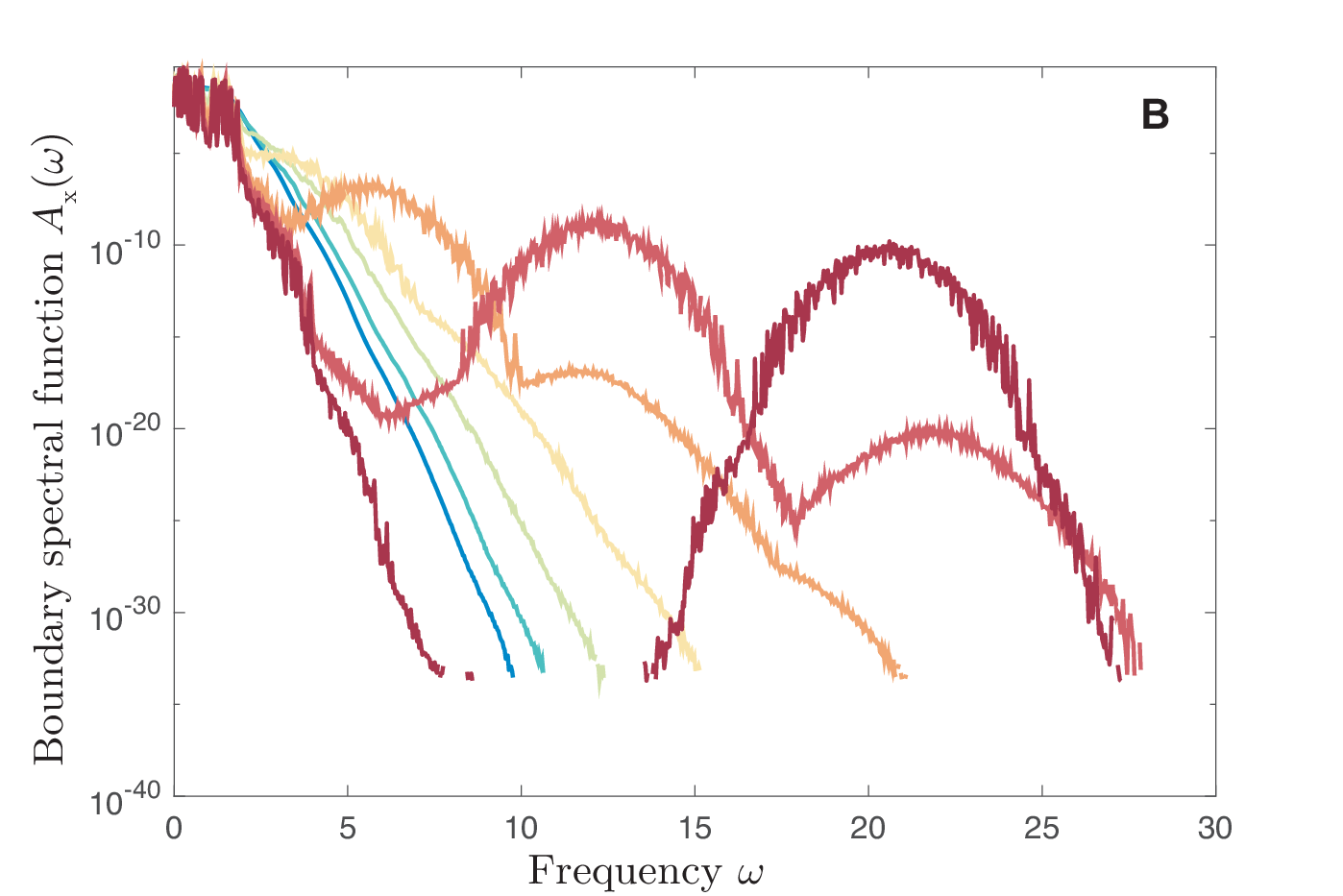}
	\includegraphics[width= 0.45\textwidth]{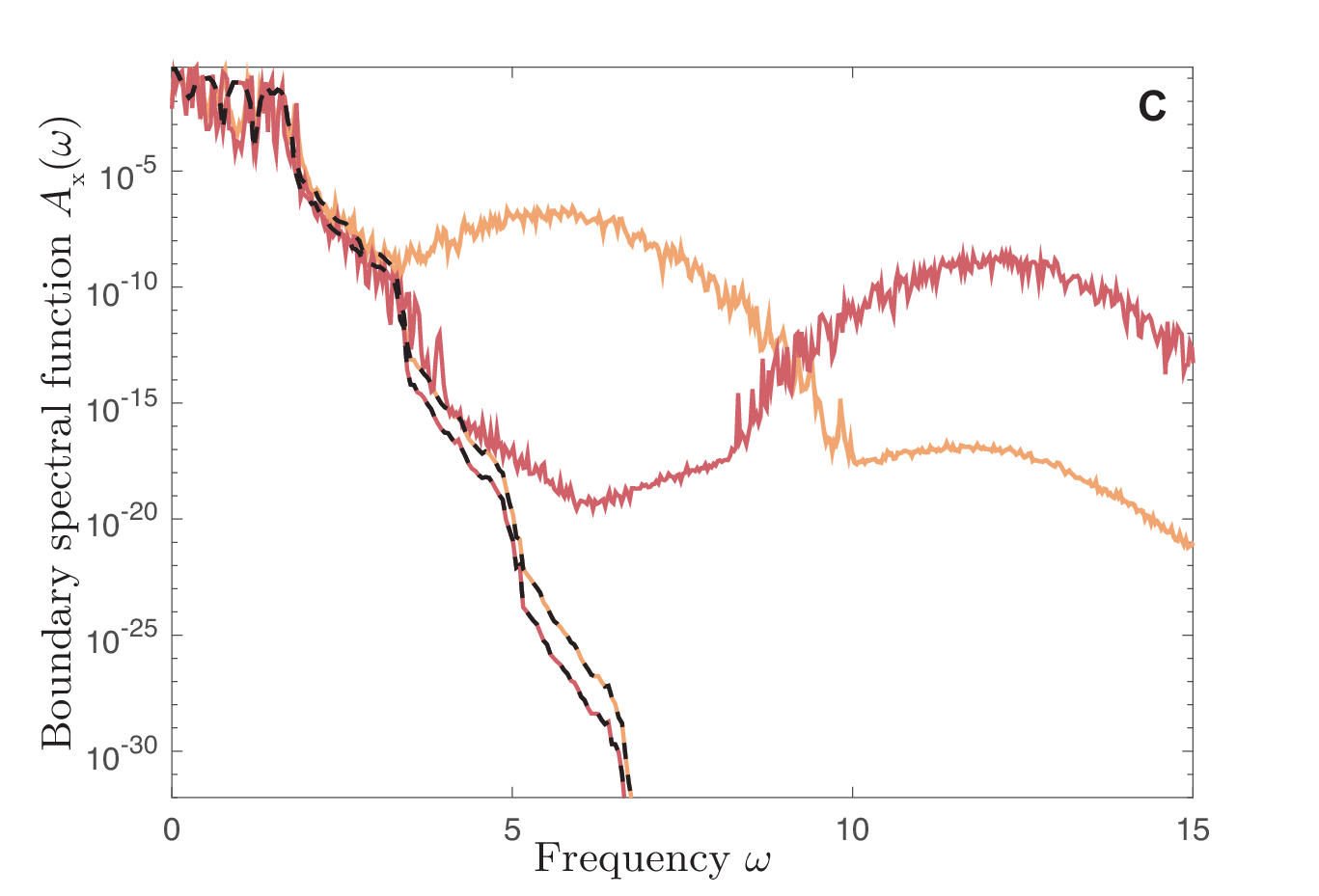}
	\caption{\textbf{Full model boundary spectral function.}  The high frequency part of the spectral function of the $S^x$ operator on the boundary of a chain with 3 impurities placed at every fourth site. Color go from blue to red with increasing $V$. Panel A shows the typical spectrum obtained by averaging $\log A_x(\omega)$ over 10 different samples and panel B shows one of those samples. Panel C shows a comparison to the effective model spectrum shown in Fig.~\ref{fig:BoundaryspecWeaklink} B for $V=40^{2/3}/2$ and $V=40^{5/6}/2$ in orange and red respectively. }
\label{fig:boundaryfullapp}
\end{figure}

In the main text we analyzed the boundary spectral function for a set of coupled blocks, such that the impurities were exactly frozen out. It is intuitively clear that freezing the impurities makes the model more non-ergodic transferring the low frequency spectral weight to high frequencies. In this Appendix we will verify that this is indeed the case. In Fig.~\ref{fig:boundaryfullapp} we show the boundary spectral function $\log(A_x(\omega))$ for the full model corresponding to the exact same parameters as the effective model spectral function shown in Fig.~\ref{fig:BoundaryspecWeaklink} (B). That is, we add 3 impurities to a chain of 15 spins. The impurity spins will cause resonances at $\omega\sim V$. In panel C we compare the spectral functions for the full and effective models for two particular values of $V:\, 10.81,5.85$. As we argued here and in the main text while $\omega \ll V$ resonances do not occur and the spectral functions of the two models are indistinguishable. However, at larger frequencies the spectral function for the full model exhibits a non-monotonic behavior due to resonant flipping of the impurity spin. Such resonant process is absent in the effective model where the other impurities are frozen and consequently its spectral function keeps monotonically decreasing with $\omega$.

\section*{Appendix F: Spectral function of the bulk spins}
In the main text, the impurity spectral function is shown to have a $1/\omega^2$ dependence at low frequency at sufficiently large $V$, where the impurity starts to decouple. In the context of MBL, similar spectral functions have been analyzed and it's been argued that they should have sub-diffusive scaling on the ergodic side leading up to the transition, i.e. $A(\omega)\sim \omega^{1-1/z}$, where $z$ goes from $z=2$ at weak disorder to $z=O(L)$ at the transition. In general, the sub diffusive behavior is attributed to Griffith's effect, where exponentially rare regions with exponentially slow transport generate anomalous transport behavior. This picture suffers from a number of problems, most notably that the same phenomenology is observed in systems with quasi-periodic potentials in which there are no rare regions. Recently, many-body resonances have been proposed as an alternative explanation~\cite{crowley2021partial}. A different phenomenological explanation of this spectral function recently emerged from the work of L. Vidmar et. al.~\cite{vidmar2021phenomenology}, which proposed a scenario of a broad distribution of the FGR relaxation rates.

\begin{figure}[h]
	\centering
	\includegraphics[width= 0.48\textwidth]{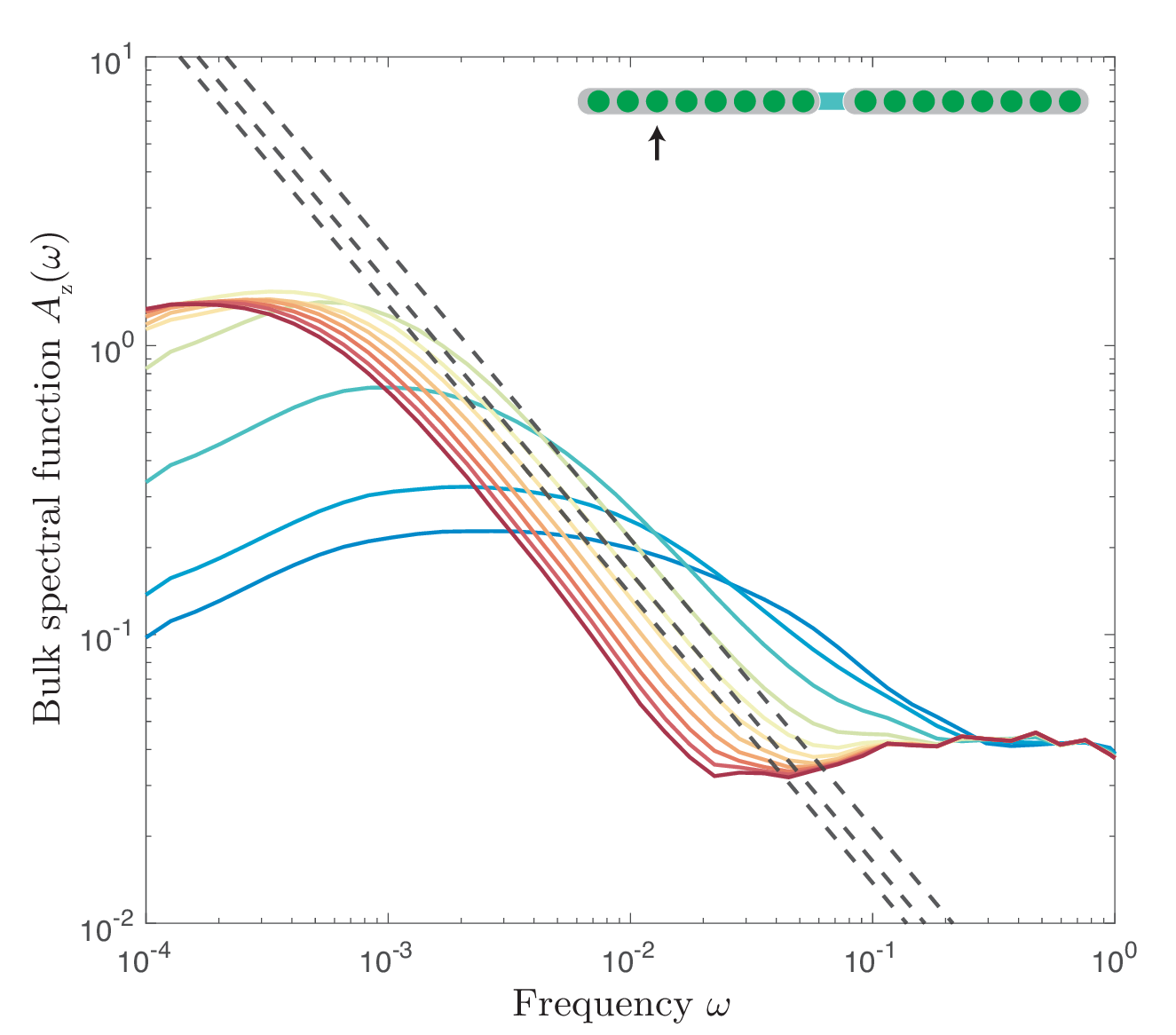}
	\caption{\textbf{Bulk spectral function.} For impurity potentials ranging from $V=1$ to $V=8$, the spectral function of a bulk spin, i.e. for $S^z_3$, is shown from blue to red in a system of $L=16$ spins described by the effective model Hamiltonian. Dashed lines are guides for the eye and indicate $1/\omega$ scaling.}
\label{fig:bulkspecapp}
\end{figure}

In Fig.~\ref{fig:bulkspecapp} we show the spectral function for a spin in the bulk of a block, i.e. for the third spin in the chain, for two weakly coupled blocks described the effective Hamiltonian~\eqref{eq:Heff}. This is precisely the same spin for which we computed the fidelity susceptibility shown in Fig.~\ref{fig:typicalAGP}. One observes a broad region where the spectral function has behavior that is close to $1/\omega$. 

\end{document}